\documentclass[11pt,a4paper]{article}
\usepackage{graphicx}
\usepackage{amsmath}
\usepackage{bm}

\def\itmb{\begin{itemize}}
\def\itme{\end{itemize}}
\def\enmb{\begin{enumerate}}
\def\enme{\end{enumerate}}
\def\eqnb{\begin{equation}}
\def\eqne{\end{equation}}

\title{On the Quadratic Phase Quaternion Domain Fourier Transform and on the Clifford algebra of $R^{3,1}$}

\author{Sadataka Furui  \\
Faculty of Science and Engineering, Teikyo University\\
2-17-12 Toyosatodai, Utsunomiya, 320-0003 Japan {\thanks
{\textit{E-mail address:} furui@umb.teikyo-u.ac.jp}}
}

\begin{document}
\maketitle
\begin{abstract}
We study application of the Clifford algebra and the Grassmann algebra to image recognitions in $(3+1)D$ using quaternions.
We construct a quaternion based wave function model with fermions and bosons of equal degrees of freedom, similar to Cartan's supersymmetric model. The Clifford algebra ${\mathcal A}_{3,1}$ is compared with ${\mathcal A}_{2,1}$ and the model applied to the $(2+1)D$ non-destructive testing is extended.

The fixed point lattice actions are calculated for 7 paths in $(3+1)D$ space with lengths less than or equal to 8 lattice units.

Comparison with the quaternion time approach of Ariel, quaternion Fourier transform of Hitzer and the tensor renormalization group approach to classical lattice models are also discussed.  
\end{abstract}

\maketitle

\section{Introduction}
Recently Hitzer\cite{Hitzer23} argued that a generalizing Castro-Minh-Tuan's "New Convolutions for Quadratic-Phase Fourier Integral Operators and their application"\cite{CMT18} in which Fourier transform is performed in $\bf R$ space can be extended to quaternion Fourier transform.
The quaternion Fourier transform was proposed by Hitzer and Sangwine \cite{HS13} and explained in the textbook \cite{Hitzer22}.  
In \cite{CMT18}, a quadratic function
\begin{equation}
Q_{(a,b,c,d,e)}(x,y):=a x^2+b x y+c y^2+d x+e y\quad a,b,c,d,e\in {\bf R}
\end{equation}
and the integral 
\begin{equation}
(Q f)(x):=\frac{1}{\sqrt{2\pi}}  \int_R e^{\sqrt{-1}Q_{(a,b,c,0,0)}(x,y) }f(y) dy
\end{equation}
was considered. Hitzer proposed that the formula can be generalized to
\begin{equation}
Q_{(a,b,c,d,e)}(x,\omega)=a x^2+b x\cdot\omega+c |\omega|^2+d\cdot x+e\cdot\omega, \quad a,b,c\in {\bf R}, \quad d,e \in {\bf H},
\end{equation}
where $\bf H$ is quaternion, and the integral
\begin{equation}
(Q h)(x) =\frac{1}{(2\pi)^2}\int_H h(x) e^{-I Q(x,\omega)}d^4\omega.
\end{equation}

In this extension of the Fourier transform of complex number space to quaternion number space have problems, 
when one considers transformations of the 3D spacial space and 1D time space. Since for (3+1) dimensional vector space $E$ and quadratic form $q$, Clifford Algebra ${\mathcal A}_{3,1}(E,q)$ is isomorphic to $M_2({\bf H})$ or biquaternion \cite{Garling11}.

Ariel\cite{Ariel21} discussed quaternion space-time in which four dimensional quaternion time
\begin{equation}
\tau^4=(i_0, {\bf i}_1\frac{x_1}{c}, {\bf i}_2\frac{x_2}{c},{\bf i}_3\frac{x_3}{c})
\end{equation}
was introduced, where $i_0,{\bf i}_1,{\bf i}_2,{\bf i}_3$ are the basis of a quaternion. The quaternion time interval is defined as
\begin{equation}
{\bf t}=t(\cos\theta, \sin\theta)=t(\sqrt{1-\frac{v^2}{c^2}},\frac{v}{c})
\end{equation}
The Lorentz time dilation was expressed by
\begin{equation}
t=|{\bf t}|=\frac{t_0}{\sqrt{1-\frac{v^2}{c^2}}},
\end{equation}
where $v$ is ${\bf x}/t$, the ratio of the space interval and the time interval.
In \cite{Ariel23}, the normed division algebra ({\bf R},{\bf C},{\bf H},{\bf O}) are postulted to be necessary tools for describing nature, and excluded biquaternions as the tools. In electrodynamics the Li\'enard-Wiechert potential contains the time delay effect $r\to r-{\bf r}\cdot{\bf u}/c$ where $\bf u$ is the velocity of an observer relative to the frame in which the charge is at rest. It is not clear whether the time delay can be expressed by a single quaternion.

Adler\cite{Adler85,Adler86} showed conditions for constructing quaternion field theory. I will review Adler's work, and try to extend the theory of propagation of 
solitonic phonon in (2+1)D Weyl fermion sea\cite{SFDS23a,SFDS23b} to quantum field theory in (4+2)D fermion sea.

The structure of this presentation is as follows. In sec.2, I discuss about Adler's quaternion Fourier transform. In sec.3, Clifford algebra ${\mathcal A}_{2,1}$ and its application and in sec.4 Clifford Algebra of ${\mathcal A}_{3,1}$ is explained. 

In sec.5 the structure of paths of (3+1)D fixed point actions used in Quantum Chromodynamics is explained. We use bosonic wave propagation in fermion sea expressed by two quaternions, and extend the method of obtaining actions adopted in (2+1)D,
in which $e_1\wedge e_2$ was interpreted as time direction. We compare the case of time shifts along $e_1e_4, e_2e_4$ and that along $e_3e_4$ which are the bases of ${\mathcal A}_{3,1}$.
 
In sec.6, a comparison with works of Hitzer, Ariel and others and a perspective of quaternion field theory is discussed.

\section{Adler's arguments on Quaternion Fourier Transformation and Hitzer's arguments}

Hitzer claimed that using a unit pure quaternion $I, (I^2=-1)$. He assumed that the unit quaternion $I$ commutes with
quaternions and derived the delta function rule
\begin{eqnarray}
&&\frac{1}{(2\pi)^4}\int_H e^{-I bx\cdot\omega}e^{+I by\cdot\omega}d^4\omega\nonumber\\
&& =\delta(b(x_r-y_r))\delta(b(x_i-y_i))\delta(b(x_j-y_j))\delta(b(x_k-y_k)),
\end{eqnarray}
where $x=x_r+x_i{\bf i}+x_j{\bf j}+x_k{\bf k}$ and $y=y_r+y_i{\bf i}+y_j{\bf j}+y_k{\bf k}$.
Products of exponential of pure quaternions is however, a complicated object. 
For $\lambda=p_0\phi_0+p_1\phi_1+p_2\phi_2+p_3\phi_3$ and $\lambda'=p_0\phi_0'+p_1\phi_1'+p_2\phi_2'+p_3\phi_3'$,
\[
\frac{1}{(2\pi)^2}\int dp e^{\sqrt{-1}(\lambda-\lambda')}=\delta(\phi-\phi')=\delta(\phi_0-\phi_0')\delta(\phi_1-\phi_1')\delta(\phi_2-\phi_2')\delta(\phi_3-\phi_3')
\]

Since $(\cos(\lambda+\lambda')+\cos(\lambda-\lambda'))/2=\cos\lambda \cos\lambda'$ and  $(-\cos(\lambda+\lambda')-\cos(\lambda-\lambda'))/2=\sin\lambda\sin\lambda'$,
the real part of $dp e^{\sqrt{-1}(\lambda-\lambda)}$ becomes $dp(\cos\lambda\cos\lambda'+\sin\lambda\sin\lambda')$.

Adler\cite{Adler86} defined for bosonic fields $\phi$,
\[
\Psi_{pe}(\phi)=\frac{1}{2\pi}\cos\lambda q_e(p)\quad \Psi_{po}(\phi)=\frac{1}{2\pi}\sin\lambda q_o(p),
\]
the completeness relation gives
\[
\int dp[\Psi_{pe}(\phi')\bar\Psi_{pe}(\phi)+\Psi_{po}(\phi')\bar\Psi_{po}(\phi)]=\delta(\phi-\phi')
\]
where a unit quaternion $q_e$ or $q_o$ depends on arbitrary ways on $p$.

The kinetic energy 
\begin{eqnarray}
H_{kin}&=&-\frac{1}{6}\sum_{i=1}^M[2e_1(\frac{\partial}{\partial\phi_1^i}\frac{\partial}{\partial\phi_3^i}-\frac{\partial}{\partial\phi_0^i}\frac{\partial}{\partial\phi_2^i})+2e_2(\frac{\partial}{\partial\phi_0^i}\frac{\partial}{\partial\phi_1^i}+\frac{\partial}{\partial\phi_2^i}\frac{\partial}{\partial\phi_3^i})\nonumber\\
&&+e_3(\frac{\partial}{\partial\phi_0^i}\frac{\partial}{\partial\phi_0^i}+\frac{\partial}{\partial\phi_3^i}\frac{\partial}{\partial\phi_3^i}
-\frac{\partial}{\partial\phi_1^i}\frac{\partial}{\partial\phi_1^i}-\frac{\partial}{\partial\phi_2^i}\frac{\partial}{\partial\phi_2^i})]
\end{eqnarray}
\[
H_{kin}\Psi_{pe}=\Psi_{pe}\lambda_e(p),\quad H_{kin}\Psi_{po}=\Psi_{po}\lambda_o(p)
\]
$\lambda_{e,o}(p)$ can be expressed as
\[
\lambda_{e,o}(p)=\frac{1}{6}(p_0^2+p_1^2+p_3^2)e_{e,o}(p)
\]
where $e_{e,o}$ is the unit pure(imaginary) quaternion, which can be given any orientation by an apropriate choice of $q_{e,o}(p)$.

 Since the generalized Quadratic Phase Quaternion Domain Fourier Transform (QPQDFT) must contain Adler's specific case, it is likely that the 2D split structure remains and the integral over $d^2p$ instead of $d^4p$ plays essential roles.

Cartan\cite{Cartan66} showed in the $\nu$ dimensional Euclidean space $E_\nu$, there exists vectors $x_1,\cdots,x_\nu$, $x_1',\cdots, x_\nu'$ and semi-spinors $\xi_0,\xi_{23},\xi_{31},\xi_{12}$ and $\xi_{123},-\xi_1,-\xi_2,-\xi_3$ and they are supersymmetric relations.

When $\nu=3$,
\begin{eqnarray}
&&x_1=\xi_{31}\xi_{12}'-\xi_{12}\xi_{31},\quad x_2=\xi_{12}\xi_{23}-\xi_{23}\xi_{12}',\quad x_3=\xi_{23}\xi_{31}'-\xi_{31}\xi_{23}'\\
&&x_1'=\xi_0\xi_{23}'-\xi_{23}\xi_0',\quad x_2'=\xi_0\xi_{31}'-\xi_{31}\xi_0',\quad x_3'=\xi_0\xi_{12}'-\xi_{12}\xi_0'.
\end{eqnarray}
and
\begin{eqnarray}
&&x_1=\xi_{123}\xi_1'-\xi_1\xi_{123},\quad x_2=\xi_{123}\xi_2'-\xi_2\xi_{123}',\quad x_3=\xi_{123}\xi_3'-\xi_3\xi_{123}'\\
&&x_1'=\xi_3\xi_2'-\xi_2\xi_3',\quad x_2'=\xi_1\xi_3'-\xi_3\xi_1',\quad x_3'=\xi_2\xi_1'-\xi_1\xi_2'
\end{eqnarray} 

When $\nu=4$, there are invariant coupling of vectors and semi-spinors
\begin{eqnarray}
\varphi ^TCX\psi&=&x^1(\xi_{12}\xi_{314}-\xi_{31}\xi_{134}-\xi_{14}\xi_{123}+\xi_{1234}\xi_1)\nonumber\\
&&+x^2(\xi_{23}\xi_{124}-\xi_{12}\xi_{234}-\xi_{34}\xi_{123}+\xi_{1234}\xi_2)\nonumber\\
&&+x^3(\xi_{31}\xi_{234}-\xi_{23}\xi_{314}-\xi_{34}\xi_{123}+\xi_{1234}\xi_3)\nonumber\\
&&+x^4(-\xi_{14}\xi_{234}-\xi_{24}\xi_{314}-\xi_{34}\xi_{124}+\xi_{1234}\xi_4)\nonumber\\
&&+x_1'(-\xi_0\xi_{234}+\xi_{23}\xi_4-\xi_{34}\xi_3+\xi_{34}\xi_2)\nonumber\\
&&+x_2'(-\xi_0\xi_{314}+\xi_{31}\xi_4-\xi_{34}\xi_1+\xi_{14}\xi_3)\nonumber\\
&&+x_3'(-\xi_0\xi_{124}+\xi_{12}\xi_4-\xi_{14}\xi_2+\xi_{24}\xi_1)\nonumber\\
&&+x_4'(\xi_0\xi_{123}-\xi_{23}\xi_1-\xi_{31}\xi_2-\xi_{12}\xi_3)
\end{eqnarray}

In the case of $\nu=4$, $x_4\to -x_4'$ transformation exists in the $G_{23}$ group.
The $G_{12}$ and $G_{13}$ groups supersymmetric transformations between $x$ and $\xi$ exist.

In performing quaternion Fourier transform, we choose discrete lattices with a certain time and a coordinate system expressed by pure quaternions. A pure quaternion $q=x e_1+y e_2 +z e_3, (x,y,z)\in R$ is chosen as $x^2+y^2+z^2=1$ and its conjugate is $\bar q=-x e_1-y e_2-z e_3$. The orthogonal 2D planes split (OPS) \cite{HS13} consists of deviding
\begin{equation}
q=q_+ +q_-, \quad q_\pm =\frac{1}{2}(q\pm e_1 q e_2). 
\end{equation}
The Hilbert space is constructed as patching the quaternion on $S^4$ around the discretized time series, and as Kodaira\cite{Kodaira92}
construct the K\"ohrer structure on the complex projected space.

\section{Clifford Algebra of (2+1)D}
In the (2+1)D Clifford action lattice simulation, the link operator of direction $e_1$, length $a$, from the position $(u_1,u_2)$ is calculated in Mathematica, by using $X=\left(\begin{array}{cc}
x&xx^-\\
I_2&x^-\end{array}\right)$, where $I_2$ is the $2\times 2$ diagonal unit matrix. The hyperplane reflection $x\to -x^-$ is represented by a multiplication of $\left(\begin{array}{cc}
0&1\\
-1&0\end{array}\right)$, a shift of length $c$ is realized by using an operator ${\mathcal T}=\left(\begin{array}{cc}
I_2&c\\
0&I_2\end{array}\right)$ as ${\mathcal T}.X.{\mathcal T}^{-1}$

The coordinate $X$ is represented by $(u_1,u_2)$ as
\begin{equation}
X(u_1,u_2)=\left(\begin{array}{cccc}
\sqrt{-1}u_1&u_2&\sqrt{-1}u_1^2-u_1u_2&u_1u_2+\sqrt{-1}u_2^2\\
-u_2&-\sqrt{-1}u_1&-u_1u_2-\sqrt{-1}u_2^2&-\sqrt{-1}u_1^2+u_1u_2\\
1&0&u_1+\sqrt{-1}u_2&0\\
0&1&0&u_1+\sqrt{-1}u_2\end{array}\right)
\end{equation}

For getting the information of anomalous scattering positions, we adopted fast Fourier Transform(FFT)  of actions alonng the $u_i$ axes $(i=1,2)$ obtained from each loops\cite{SFDS23a}.

In signal processing\cite{MFLBCB23}, there is a Fourier transform method using the Riesz wavelet transform\cite{USVDV09}. The wavelet transform is different from FFT in taking into account the shape of amplitudes squared. The Riesz transform in 2D is expressed as first calculate
\begin{equation}
{\bf f}_R({\bf x})=\left(\begin{array}{c}
f_1({\bf x})\\
f_2({\bf x})\end{array}\right)=\left(\begin{array}{c}
h_x*f({\bf x})\\
h_y*f({\bf x})\end{array}\right)
\end{equation}
where ${\bf x}=(x,y)$ is the input signal, $h_x({\bf x})=x/(2\pi ||{\bf x}||^3)$, $h_y({\bf x})=y/(2\pi||{\bf x}||^3)$\cite{USVDV09}.

In the TR-NEWS, convolution of signals with windows functions, which are different from $h_x$ and $h_y$ are used, and we adopt the short-time Fourier transform (STFT)\cite{wiki22}.
\begin{equation}
{\rm STFT}\{x(t)\}(\tau,\omega)\equiv X(\tau,\omega)=\int_{-\infty}^\infty x(t) w(t-\tau)e^{-\sqrt{-1}\omega t} dt
\end{equation}
The discrete time STFT is
\begin{equation}
{\rm STFT}\{x[n]\}(m,\omega)=\sum_{n=-\infty}^\infty x[n]w[n-m]e^{-\sqrt{-1}\omega n}
\end{equation}

Inverse STFT is
\begin{equation}
x(t)w(t-\tau)=\frac{1}{2\pi}\int_{-\infty}^\infty X(\tau,\omega) e^{+\sqrt{-1}\omega t} d\omega.
\end{equation}

An extension to $(3+1)D$ will be discussed in sec.6.

\section{Clifford Algebra of (3+1)D}
A mapping $R^3\to R^{3,1}$ is obtained by taking $u_1=x, u_2=y ,u_3=z$, and a vector $X\in R^{3,1}$ is expressed as 
\begin{equation}
X=x e_1+y e_2+z e_3+t e_4,\\
\end{equation}

In $R^{3,1}$, the basis are $4\times 4$ matrices\cite{Lounesto01}
\begin{eqnarray}
&&e_1=\left(\begin{array}{cccc}
1&0&0&0\\
0&-1&0&0\\
0&0&-1&0\\
0&0&0&1\end{array}\right),\quad e_2=\left(\begin{array}{cccc}
0&1&0&0\\
1&0&0&0\\
0&0&0&1\\
0&0&1&0\end{array}\right),\nonumber\\
&&e_3=\left(\begin{array}{cccc}
0&0&1&0\\
0&0&0&-1\\
1&0&0&0\\
0&-1&0&0\end{array}\right),\quad e_4=\left(\begin{array}{cccc}
0&-1&0&0\\
1&0&0&0\\
0&0&0&-1\\
0&0&1&0\end{array}\right).
\end{eqnarray}

 $e_1=-e_2e_3, e_2=-e_3e_1, e_3=-e_1e_2$ are the bases of $Cl^+_3$\cite{Lounesto01}, or ${\mathcal A}_3^+$\cite{Garling11}.
The sum of the products $e_i{\bar e}_i+{\bar e}_i e_i$  is 0 due to anti-commutativity of quaternions.

In analogy to the $(2+1)D$ case, one could map $x,y,z,t$ on $S^4$ as
\begin{equation}
X=\frac{2u_1}{1+|u|^2}e_1+\frac{2u_2}{1+|u|^2}e_2+\frac{2u_3}{1+|u|^2}e_3+\frac{1-|u|^2}{1+|u|^2}e_4,
\end{equation}
where $|u|^2=u_1^2+u_2^2+u_3^2$.

Hitzer and Sangwine's\cite{HS13} OPS of quaternion space suggests that not only $e_4, e_3$, but also $e_4, e_1$ and $e_4, e_2$
can be chosen as the split bases.

We define a reflection matrix ${\rm ref}=\left(\begin{array}{cccc}
0&0&1&0\\
0&0&0&-1\\
1&0&0&0\\
0&-1&0&0\end{array}\right)$
and define ${\bar e_i}=-{\rm ref}.e_i$\\ ($i=1,\cdots ,4$). For $i=1,2,3$ $e_i.\bar e_i=\left(\begin{array}{cc}
0&I_2\\
-I_2&0\end{array}\right)$ and $e_4.{\bar e_4}=\left(\begin{array}{cc}
0&-I_2\\
I_2&0\end{array}\right)$. 

The sign difference of $e_4.\bar e_4$ matches the Minkowski's metric.

Following the conformal treatment of Clifford algebra\cite{Porteous95}, we define
\begin{equation}
{\mathcal X}=\left(\begin{array}{cc}
X&X{\bar X}\\
I_4&{\bar X}\end{array}\right)
\end{equation}
where $I_4$ is 4 dimensional diagonal matrix.  When there is one quaternion, the method of sandwitching $\mathcal X$ between extended Vahlen matrices \cite{Vahlen02}, adopted in $(2+1)D$ works.

Lounesto\cite{Lounesto01} and Vaz\cite{Vaz16} defined a vector in $Cl_3\simeq M_2({\bf C})$ as
\begin{equation}
u=u_0+u_1 e_1+u_2 e_2+u_3 e_3+u_{12}e_{12}+u_{13}e_{13}+u_{23}e_{23}+u_{123}e_{123}
\end{equation}
and for an even element $u\in Cl_3^+$ and 4 dimensional base $(f_1,f_2,f_3,f_4)$, expressed
\begin{eqnarray}
u\psi&=&(u_0+u_1 e_{23}+u_2 e_{31}+u_3 e_{23})(\psi_0 f_0+\psi_1 f_1+\psi_2 f_2+\psi_3 f_3)\nonumber\\
&\simeq&\left(\begin{array}{cccc}
u_0&-u_1&-u_2&-u_3\\
u_1&u_0&u_3&-u_2\\
u_2&-u_3&u_0&u_1\\
u_3&u_2&-u_1&u_0\end{array}\right)\left(\begin{array}{c}
\psi_0\\
\psi_1\\
\psi_2\\
\psi_3\end{array}\right)
\end{eqnarray}
and showed $Cl_3^+\simeq {\bf H}$. Vaz described the electron's Dirac equation in the form
\begin{equation}
\sqrt{-1}\hbar(\partial_t+\alpha^k\partial_k)\psi=m\beta(\psi),
\end{equation}
where
\begin{eqnarray}
&&\beta(\alpha^i\psi)+\alpha^i\beta(\psi)=0,\quad i=1,2,3\\
&&(\sqrt{-1}\beta)^2(\psi)=\sqrt{-1}\beta(\sqrt{-1}\beta(\psi))=-\psi.
\end{eqnarray}
The author claimed by a proper choice of $\beta$, $\alpha^k$ can be chosen as the Pauli matrices.

There are discussions on the $\beta$ in more general framework by Hestenes\cite{Hestenes20}. 

The algebra in $R^{3,1}$ is not isomorphic to one quaternion. The algebra ${\mathcal A}_{3,1}$ is isomolphic to $M_2({\bf H})$.

Garling \cite{Garling11} performed the mapping $\tilde\gamma=R^{3,1}\to M_2({\bf C})$ as Dyson\cite{Dyson85}
\begin{equation}
x=\lambda_0 I+\sum_{1\le i<j\le 4}\lambda_{ij}e_i e_j+\lambda_\Omega e_\Omega
\end{equation}
and set $p=\lambda_0+\sqrt{-1}\lambda_\Omega, q=-(\lambda_{13}+\sqrt{-1}\lambda_{34}),$ $r=-(\lambda_{14}+\sqrt{-1}\lambda_{23}),
s=-(\lambda_{34}+\sqrt{-1}\lambda_{12})$ for the mapping equivalent to $M_2({\bf C})$.

Here
\begin{equation}
e_2 e_3=\left(\begin{array}{cccc}
0&-\sqrt{-1}&0&0\\
-\sqrt{-1}&0&0&0\\
0&0&0&\sqrt{-1}\\
0&0&\sqrt{-1}&0\end{array}\right),\quad
e_1 e_3=\left(\begin{array}{cccc}
0&1&0&0\\
-1&0&0&0\\
0&0&0&1\\
0&0&-1&0\end{array}\right)
\end{equation}

\begin{equation}
e_1 e_2=\left(\begin{array}{cccc}
-\sqrt{-1}&0&0&0\\
0&\sqrt{-1}&0&0\\
0&0&-\sqrt{-1}&0\\
0&0&0&\sqrt{-1}\end{array}\right),\quad
e_1 e_4=\left(\begin{array}{cccc}
0&-1&0&0\\
-1&0&0&0\\
0&0&0&1\\
0&0&1&0\end{array}\right)
\end{equation}
\begin{equation}
e_2 e_4=\left(\begin{array}{cccc}
0&\sqrt{-1}&0&0\\
-\sqrt{-1}&0&0&0\\
0&0&0&\sqrt{-1}\\
0&0&-\sqrt{-1}&0\end{array}\right),\quad
e_3 e_4=\left(\begin{array}{cccc}
-1&0&0&0\\
0&1&0&0\\
0&0&1&0\\
0&0&0&-1\end{array}\right)
\end{equation}

For $x=\lambda_0 I+\sum_{1\leq i<j\leq 4}\lambda_{ij}e_i e_j $, and the Dirac matrices described by 
\begin{equation}
Q=\left(\begin{array}{cc}
0&1\\
1&0\end{array}\right), J=\left(\begin{array}{cc}
0&-1\\
1&0\end{array}\right), U=\left(\begin{array}{cc}
1&0\\
0&-1\end{array}\right)
\end{equation}
as
\begin{eqnarray}
&&\gamma(e_1 e_3)=-J\otimes I=\left(\begin{array}{cccc}
0&0&1&0\\
0&0&0&1\\
-1&0&0&0\\
0&-1&0&0\end{array}\right)\\
&&\gamma(e_2 e_4)=-\sqrt{-1}J\otimes U=\left(\begin{array}{cccc}
0&0&\sqrt{-1}&0\\
0&0&0&-\sqrt{-1}\\
-\sqrt{-1}&0&0&0\\
0&\sqrt{-1}&0&0\end{array}\right)\\
&&\gamma(e_1 e_4)=-Q\otimes U=\left(\begin{array}{cccc}
0&0&-1&0\\
0&0&0&1\\
-1&0&0&0\\
0&1&0&0\end{array}\right)
\end{eqnarray}
\begin{eqnarray}
&&\gamma(e_2 e_3)=-\sqrt{-1}Q\otimes I=\left(\begin{array}{cccc}
0&0&-\sqrt{-1}&0\\
0&0&0&-\sqrt{-1}\\
-\sqrt{-1}&0&0&0\\
0&-\sqrt{-1}&0&0\end{array}\right)
\end{eqnarray}
\begin{eqnarray}
&&\gamma(e_3 e_4)=-U\otimes U=\left(\begin{array}{cccc}
-1&0&0&0\\
0&1&0&0\\
0&0&1&0\\
0&0&0&-1\end{array}\right)\\
&&\gamma(e_1 e_2)=-\sqrt{-1}U\otimes I=\left(\begin{array}{cccc}
-\sqrt{-1}&0&0&0\\
0&-\sqrt{-1}&0&0\\
0&0&\sqrt{-1}&0\\
0&0&0&\sqrt{-1}\end{array}\right)
\end{eqnarray}

\begin{equation}
\gamma(x)-\lambda_0 I_4=\left(\begin{array}{cccc}
-t&0&u-v&0\\
0&\bar t&0&\bar u+\bar v\\
-u-v&0&t&0\\
0&-\bar u+\bar v&0&-\bar t\end{array}\right)
\end{equation}
where $t=\lambda_{34}+\sqrt{-1}\lambda_{12}$, $u=\lambda_{13}+\sqrt{-1}\lambda_{24}$, $v=\lambda_{14}+\sqrt{-1}\lambda_{23}$. 

The reflections of  $e_i e_j$are denoted as $\overline{e_i e_j}$.
\begin{equation}
\overline{e_2 e_3}=\left(\begin{array}{cccc}
0&0&0&\sqrt{-1}\\
0&0&-\sqrt{-1}&0\\
0&-\sqrt{-1}&0&0\\
\sqrt{-1}&0&0&0\end{array}\right),\quad
\overline{e_1 e_3}=\left(\begin{array}{cccc}
0&0&0&1\\
0&0&1&0\\
0&1&0&0\\
1&0&0&0\end{array}\right)
\end{equation}

\begin{equation}
\overline{e_1 e_2}=\left(\begin{array}{cccc}
0&0&-\sqrt{-1}&0\\
0&0&0&-\sqrt{-1}\\
-\sqrt{-1}&0&0&0\\
0&-\sqrt{-1}&0&0\end{array}\right),\quad
\overline{e_1 e_4}=\left(\begin{array}{cccc}
0&0&0&1\\
0&0&-1&0\\
0&-1&0&0\\
1&0&0&0\end{array}\right)
\end{equation}
\begin{equation}
\overline{e_2 e_4}=\left(\begin{array}{cccc}
0&0&0&\sqrt{-1}\\
0&0&\sqrt{-1}&0\\
0&\sqrt{-1}&0&0\\
\sqrt{-1}&0&0&0\end{array}\right),\quad
\overline{e_3 e_4}=\left(\begin{array}{cccc}
0&0&1&0\\
0&0&0&1\\
-1&0&0&0\\
0&-1&0&0\end{array}\right).
\end{equation}

The product of the multiple and its reflection is
\begin{eqnarray}
&&e_1 e_2.\overline{e_1 e_2}=-e_1 e_3.\overline{e_1 e_3}=e_1e_4.\overline{e_1e_4}=-e_2e_3.\overline{e_2 e_3}=
e_2e_4.\overline{e_2 e_4}=-e_3 e_4.\overline{e_3e_4}\nonumber\\
&&=\left(\begin{array}{cccc}
0&0&-1&0\\
0&0&0&1\\
-1&0&0&0\\
0&1&0&0\end{array}\right)
\end{eqnarray} 

The shift transformation becomes $\mathcal TX{\mathcal T}^{-1}=X'$, where
${\mathcal T}=\left(\begin{array}{cc}
I_4&c\\
0&I_4\end{array}\right)$, and modify $c$ suggested by Garling\cite{Garling11} 
\begin{equation}
c=c_{23}e_2e_3+c_{13}e_1e_3+c_{12}e_1e_2+c_{14}e_1 e_4+c_{24}e_2e_4+c_{34}e_3 e_4+c_0 I+c_\Omega e_\Omega,
\end{equation}
for calculation of actions in (3+1)D, this 8 dimensional shift vector is not suitable. 

Therefore we replace the base $e_1$ by $e_2e_3$, $e_2$ by $e_3 e_1$, $e_3$ by $e_1 e_2$. The $e_4$ is replaced by $e_1e_4, e_2e_4$ or $e_3e_4$.

Therefore we choose shift vectors in (3+1)D as
\begin{eqnarray}
c_1&=&c_{23}e_2e_3+c_{13}e_1e_3+c_{12}e_1e_2+c_{14}e_1 e_4\nonumber\\
c_2&=&c_{23}e_2e_3+c_{13}e_1e_3+c_{12}e_1e_2+c_{24}e_2 e_4\nonumber\\
c_3&=&c_{23}e_2e_3+c_{13}e_1e_3+c_{12}e_1e_2+c_{34}e_3 e_4
\end{eqnarray}
and corresponding momentum space position vectors
\begin{eqnarray}
x_1&=&x e_2e_3+y e_1e_3+z e_1e_2+t e_1 e_4\nonumber\\
x_2&=&x e_2e_3+y e_1e_3+z e_1e_2+t e_2 e_4\nonumber\\
x_3&=&x e_2e_3+y e_1e_3+z e_1e_2+t e_3 e_4.
\end{eqnarray}
When the time component is $t e_ie_4$ and $t\overline{e_ie_4}$, $(i=1,2,3)$, we define
\begin{equation}
\Phi_i=\left(\begin{array}{cc}
X_i&X_i\bar X_i\\
I_4&\bar X_i\end{array}\right), 
\end{equation}
where $I_4$ is a 4 dimensional diagonal matrix, and $X_i,\bar X_i$ are $4\times 4$ matrices.

When the time component is $t e_1e_4$ and $t\overline{e_1e_4}$,
\begin{equation}
X_1=\left(\begin{array}{cccc}
-\sqrt{-1}z&-t-\zeta&0&0\\
-t+\bar\zeta&\sqrt{-1}z&0&0\\
0&0&-\sqrt{-1}z&t-\bar\zeta\\
0&0&t+\zeta&\sqrt{-1}z\end{array}\right),
\bar X_1=\left(\begin{array}{cccc}
0&0&-\sqrt{-1}z&t-\bar\zeta\\
0&0&-t-\zeta&-\sqrt{-1}z\\
-\sqrt{-1}z&-t-\zeta&0&0\\
t-\bar\zeta&-\sqrt{-1}z&0&0\end{array}\right)\\
\end{equation}
where $\zeta=\sqrt{-1}x-y, \bar\zeta=-\sqrt{-1}x-y$.
\begin{equation}
X_1\bar X_1=\left(\begin{array}{cccc}
0&0&(t+\zeta)^2-z^2&-2\sqrt{-1}yz\\
0&0&2\sqrt{-1}yz&-(t-\bar\zeta)^2+z^2\\
(t-\bar\zeta)^2-z^2&-2\sqrt{-1}yz&0&0\\
2\sqrt{-1}yz&-(t+\zeta)^2+z^2&0&0\end{array}\right)
\end{equation}

When the time component is $t e_2e_4$ and $t\overline{e_2e_4}$,
\begin{equation}
X_2=\left(\begin{array}{cccc}
-\sqrt{-1}z&\sqrt{-1}t-\zeta&0&0\\
\sqrt{-1}t+\bar\zeta&\sqrt{-1}z&0&0\\
0&0&-\sqrt{-}z&\sqrt{-1}t-\bar\zeta\\
0&0&-\sqrt{-1}t-\bar\zeta&\sqrt{-1}z\end{array}\right)
\end{equation}
\begin{equation}
\bar X_2=\left(\begin{array}{cccc}
0&0&-\sqrt{-1}z&\sqrt{-1}t-\bar\zeta\\
0&0&\sqrt{-1}t-\zeta&-\sqrt{-1}z\\
-\sqrt{-1}z&\sqrt{-1}t-\zeta&0&0\\
\sqrt{-1}t-\bar\zeta&-\sqrt{-1}z&0&0\end{array}\right).
\end{equation}
\begin{equation}
X_2\bar X_2=\left(\begin{array}{cccc}
 0&0&(\sqrt{-1}t-\zeta)^2-z^2&2(t-\sqrt{-1}y)z\\
0&0&-2(t-\sqrt{-1}y)z&-(\sqrt{-1}t-\bar\zeta)^2+z^2\\
(\sqrt{-1}t-\zeta)^2-z^2&2(t-\sqrt{-1}y)z&0&0\\
-2(t-\sqrt{-1}y)z&-(\sqrt{-1}t-\zeta)^2+z^2&0&0\end{array}\right)
\end{equation}
When the time component is $t e_3e_4$ and $t\overline{e_3e_4}$,
\begin{equation}
X_3=\left(\begin{array}{cccc}
-t-\sqrt{-1}z&-\zeta&0&0\\
\bar\zeta&t+\sqrt{-1}z&0&0\\
0&0&t-\sqrt{-1}z&-\bar\zeta\\
0&0&\zeta&-t+\sqrt{-1}z\end{array}\right)
\end{equation}
\begin{equation}
\bar X_3=\left(\begin{array}{cccc}
0&0&t-\sqrt{-1}z&-\bar\zeta\\
0&0&-\zeta&t-\sqrt{-1}z\\
-t-\sqrt{-1}z&-\zeta&0&0\\
-\bar\zeta&-t-\sqrt{-1}z&0&0\end{array}\right).
\end{equation}
\begin{equation}
X_3\bar X_3=\left(\begin{array}{cccc}
0&0&\zeta^2-t^2-z^2&-2\sqrt{-1}(tx+yz)\\
0&0&-2\sqrt{-1}(tx-yz)&-\zeta^2+t^2+z^2\\
 \zeta^2-t^2-z^2&-2\sqrt{-1}(tx+yz)&0&0\\
-2\sqrt{-1}(tx-yz)&-\zeta^2+t^2+z^2&0&0\end{array}\right)
\end{equation}



Dyson pointed out that for a quaternion \cite{Dyson85}
\begin{equation}
q=\left(\begin{array}{cc}
a&b\\
c&d\end{array}\right),\quad \bar q=\left(\begin{array}{cc}
d&-b\\
-c&a\end{array}\right), \quad
\tau_3=\left(\begin{array}{cc}
\sqrt{-1}&0\\
0&-\sqrt{-1}\end{array}\right), 
\end{equation}
\begin{equation}
\bar q  \tau_3 q=\sqrt{-1}\left(\begin{array}{cc}
bc+ad&2bd\\
-2ac&-bc -ad\end{array}\right)
\end{equation}
becomes a pure imaginary matrix.

In the present work, we do not consider gauge actions and consider propagation of Bosonic solitons in fermion sea.

In lattice simulations we consider paths that start from a point in space-time and return to the starting point which contain time axes.
The continuation of a  path of direction $e_i e_j$ and $e_k e_l$ is chosen such that either $i=k$ or $l$, or either $j=k$ or $l$ and the time axis is chosen one of $e_1 e_4$, $e_2 e_4$ and $e_3 e_4$. The step 8 returns to the step 0 cyclically. 

\section{Fixed Point actions in (3+1)D}
The Fixed Point (FP) actions are characterized by spacial lattice unit $\Delta_u=\frac{1}{2^{n+2}}$ and time step unit $\delta$. The loops starts from $(x,y,z,t)=(i,j,k,0)$, and returns to the initial position. Loops of DeGrand et al.\cite{DGHHN95} that contain links of both $z$ and $t$ are $L19,L20,L21,L22,L23,L24,L25$. 

The paths $L19-25$ considered by DeGrand et al\cite{DGHHN95} are shown in Fig. 1,2,3,4,5,6 and Fig.7. A pair of balls at edges is the position where time shifts (hysteresis effects) occur.

\begin{figure}[htb]
\begin{minipage}{0.47\linewidth}
\begin{center}
\includegraphics[width=4cm,angle=0,clip]{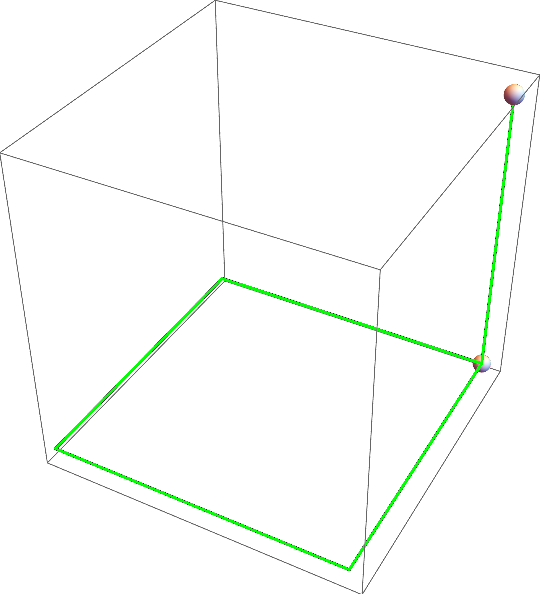}
\end{center}
\caption{ L19. The upper right corner ball is $e_2e_4$, the lower right corner one is $-e_2e_4$}
\end{minipage}
\quad
\begin{minipage}{0.47\linewidth}
\begin{center}
\includegraphics[width=4cm,angle=0,clip]{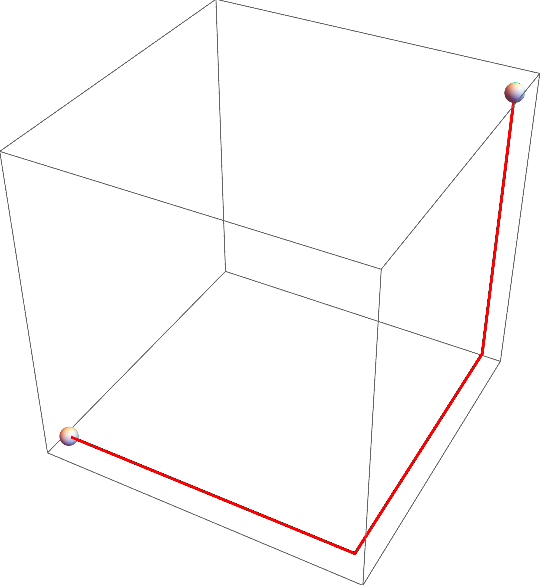}
\end{center}
\caption{ L20. The upper right corner ball is $e_2e_4$, the lower left corner one is $-e_2e_4$}
\end{minipage}
\end{figure}
\begin{figure*}[htb]
\begin{minipage}{0.47\linewidth}
\begin{center}
\includegraphics[width=4cm,angle=0,clip]{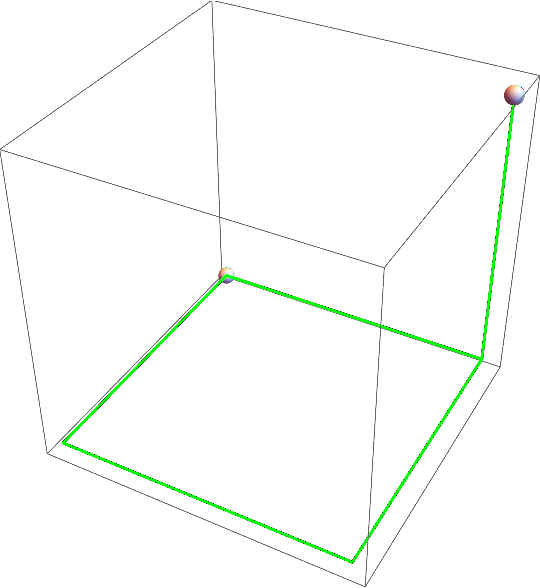}
\end{center}
\caption{ L21. The upper right corner ball is $e_1e_4/e_2e_4$, the lower left corner one is $-e_3e_4$. }
\end{minipage}
\quad
\begin{minipage}{0.47\linewidth}
\begin{center}
\includegraphics[width=4cm,angle=0,clip]{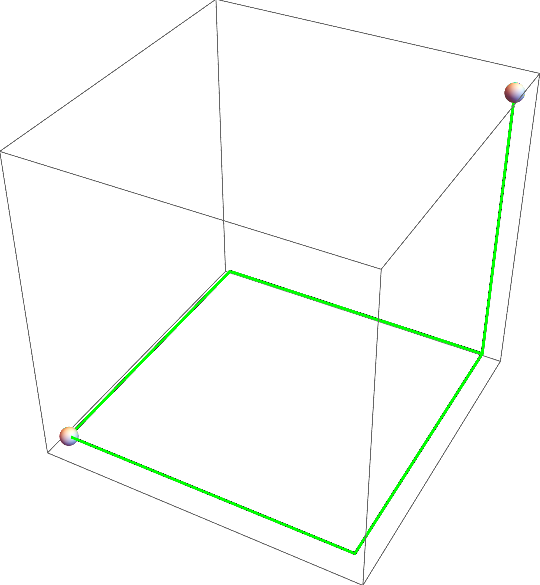}
\end{center}
\caption{ L22. The upper right corner ball is $e_1e_4/e_2e_4$, the lower left corner one is $-e_3e_4$.}
\end{minipage}
\end{figure*}
\begin{figure*}[htb]
\begin{minipage}{0.47\linewidth}
\begin{center}
\includegraphics[width=4cm,angle=0,clip]{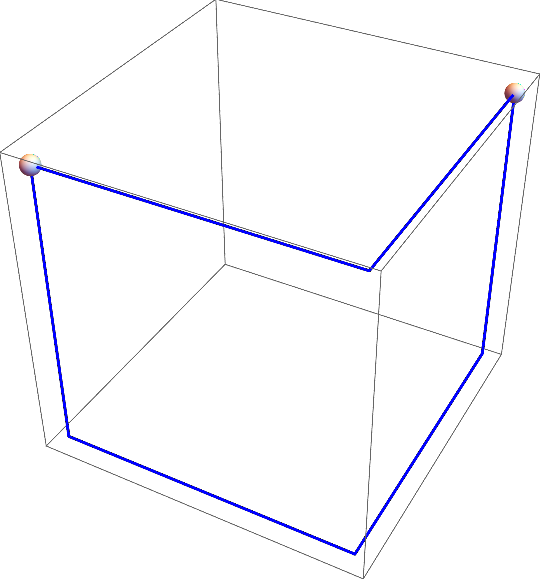}
\end{center}
\caption{ L23. The upper right corner ball is $e_1e_4$, the upper left corner one is $-e_2e_4$.}
\end{minipage}
\quad
\begin{minipage}{0.47\linewidth}
\begin{center}
\includegraphics[width=4cm,angle=0,clip]{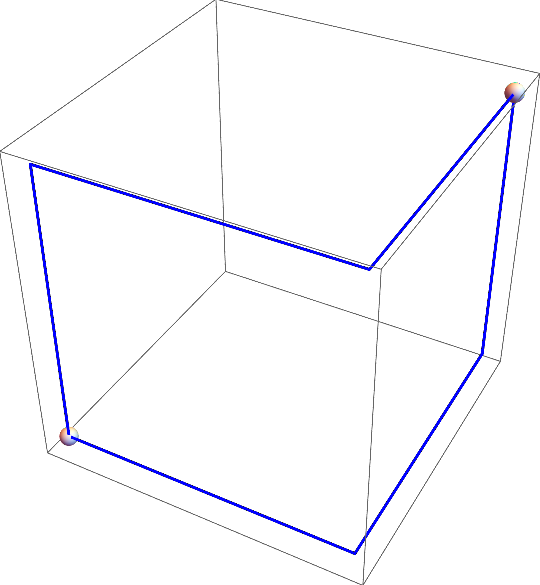}
\end{center}
\caption{ L24. The upper right corner ball is $e_1e_4$, the lower left corner one is $-e_2e_4$}
\end{minipage}
\end{figure*}
\begin{figure}[htb]
\begin{center}
\includegraphics[width=4cm,angle=0,clip]{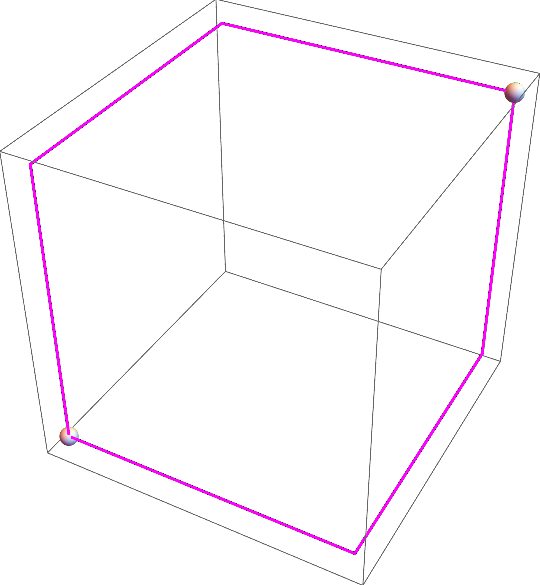}
\end{center}
\caption{ L25. The upper right corner ball is $e_2e_4$, the lower left corner one is $-e_2e_4$}
\end{figure}

The 7 FP actions that contain paths along the $t$ axis are shown in the Table 1. there are ambiguity in the choice of $e_1e_4$ or $e_2e_4$, but we choose one if there is a partner of the opposite time direction, 

Since $e_1 e_4$, $e_2 e_4$,$e_3 e_4$ have different space-time transformations, when the partner of $e_k e_4$ is $-e_j e_4$ $(j\ne k)$, nontrivial transformations appear.  

In the case of $L19$, the step 5 is fixed to be $-e_2e_4$, and the step 3 is fixed to be $e_2e_4$. Similarly in $L25$, $\pm e_2e_4$ pair appear. 

In the case of $L20$,the step 7 is $-e_2e_4$ and the step 3 is fixed to be $e_2e_4$.

In the case of $L21$ and $L22$, both $e_1e_2$ or $e_2e_1$ partner is $-e_3e_4$.

In the case of $L23$ and $L24$, $e_1e_4, -e_2e_4$ pair appear.

In the table 1, among 7 $\pm e_ie_4$ pairs there are 2 $-e_3e_4$ and 12 $ e_1e_4$ or $e_2e_4$ pairs.

\begin{tabular}{r|cccccccc}
 step & 0&1&2&3 &4&5&6&7\\
\hline
L19&x&y&z&t&-z&-t&-x&-y\\
&23&31&12&24&-12&-24&-23&-31\\
\hline
L20&x&y&z&t&-z&-y&-x&-t\\
&23&31&12&24&-12&-31&-23&-24\\
\hline
L25&x&y&z&t&-x&-y&-z&-t\\
&23&31&12&24&-23&-13&-12&-24\\
\end{tabular}

 Table 1. Directions of the wave front of loops L19, L20,  L25 
\begin{itemize}
\item L21

{\small
\begin{tabular}{cccccccccccccccc}
\hline
x&y&z&t&-z&-x&-t&-y&-x&-y&-z&-t&z&x&t&y\\
23&31&12&14/24&-12&-23&-34&-13&-23&-31&-12&-14/24&12&23&34&13\\
\end{tabular}
}

\item L22

{\small
\begin{tabular}{cccccccccccccccc}
\hline
x&y&z&t&-z&-x&-y&-t&-x&-y&-z&-t&z&x&y&t\\
23&31&12&14/24&-12&-23&-31&-34&-23&-31&-12&-14/24&12&23&31&34\\
\end{tabular}
}

\item L23

{\small
\begin{tabular}{cccccccccccccccc}
\hline
x&y&z&t&-y&-x&-t&-z&-x&-y&-z&-t&y&x&t&z\\
23&31&12&14&-31&-23&-24&-12&-23&-31&-12&-14&31&23&24&12\\
\end{tabular}
}

\item L24

{\small
\begin{tabular}{cccccccccccccccc}
\hline
x&y&z&t&-y&-x&-z&-t&-x&-y&-z&-t&y&x&z&t\\
23&31&12&14&-31&-23&-12&-24&-23&-31&-12&-14&31&23&12&24\\
\end{tabular}
}

\end{itemize}
 Table 2. Directions of the wave front of loops in the $e_1,e_2,e_3,e_4$ bases (the first line) and in the $e_ie_j$ bases (the second line) of $L21, L22, L23$ and $L24$.
\vskip 0.5 cm
The first line of each loop indicates the direction of the wave front. 
The second line indicates $ij$ of the basis $e_ie_j$ of the $R^{3,1}\simeq M_2({\bf H})$. The transformations are chosen such that $i$ or $j$ of subsequent $e_i e_j$ is equal, and that $e_i e_4$ and $-e_i e_4$ appear in the sequences.  

The path of  $L19,L20,L21,L22$ are shown in Fig.1, 2, 3, 4 respectively.

 The path of $L23,L24,L25$ are shown in Fig.5, 6, 7 respectively.

In the calculation of 7 loop actions, we choose 3 random numbers $(u_1.u_2,u_3)$ each in $(0,1)$
and calculate $(s_1,s_2,s_3,s_4)$ in $(0,\Delta_u)$. 
Then we choose 8 random numbers $(a_1,a_2,b_1,b_2,c_1,c_2,d_1.d_2)$ each in $(0,1)$ of arbitrary number of sets. As a test I chosse 4 sets.

\begin{itemize}
\item 1) Define $\tilde j_1=j+s_1, x_1=\tilde j_1 e_2e_3$, $\bar x_1=\tilde j_1\overline{e_2e_3}$ and $X_1=\left(\begin{array}{cc} 
 x_1 ,& x_1\bar x_1\\
I_4& \bar x_1\end{array}\right)$.

Define $V_1=\left(\begin{array}{cc}
 a_1 I_4& b_1 e_2e_3\\
 c_1 e_2e_3&0\end{array}\right)$ and $V^\dagger_1=\left(\begin{array}{cc}
0&c_1\overline{e_2e_3}\\
b_1\overline{e_2e_3}&a_1\bar I_4\end{array}\right)$

Calculate $V_1 X_1 V^\dagger_1-X_1$.

\item 2) Define $\tilde j_2=j_0+s_2, x_2=\tilde j_1 e_2e_3+\tilde j_2 e_1e_3$, $\bar x_2=\tilde j_1 \overline{e_2e_3}+\tilde j_2\overline{e_1e_3}$, where $j_0$ is a fixed number in order to see the dependence of action as a function of $j$.

Define $X_2=\left(\begin{array}{cc} 
 x_2 ,& x_2\bar x_2\\
I_4& \bar x_2\end{array}\right)$,  $V_2=\left(\begin{array}{cc}
 a_1 I_4+a_2 e_1e_3& 0\\
 0&d_2 e_1e_3\end{array}\right)$ and $V^\dagger_1=\left(\begin{array}{cc}
d_2\overline{e_1e_3}&0\\
0&a_1\bar I_4+a_2\overline{e_1e_3}\end{array}\right)$.

Calculate $V_2 X_2 V^\dagger_2-X_2$.

\item 3) Define $\tilde j_3=j_0+s_3, x_3=\tilde j_1 e_2e_3+\tilde j_2 e_1e_3+\tilde j_3 e_1e_2$, $\bar x_3=\tilde j_1 \overline{e_1e_2}+\tilde j_2\overline{e_1e_2}+\tilde j_3\overline{e_1e_2}$.  

Define $X_3=\left(\begin{array}{cc} 
 x_3 ,& x_3\bar x_3\\
I_4& \bar x_3\end{array}\right)$, $V_3=\left(\begin{array}{cc}
 a_1 I_4& 0\\
 0&d_1 e_2e_4\end{array}\right)$ and $V^\dagger_3=\left(\begin{array}{cc}
d_1\overline{e_2e_4}&0\\
0&a_1\bar I_4\end{array}\right)$.

Calculate $V_3 X_3 V^\dagger_3-X_3$.

 From step 4, the simulation depends on paths. In order to check my program, I first replace ambiguous $e_1e_4$ and $e_2e_4$ in $L21$ and $L22$ by $e_3e_4$.
To check my program, I also replace  $e_1e_4$ and $-e_2e_4$ in $L23$ and $L24$ by $e_3e_4$ and $-e_3e_4$, respectively.
\end{itemize}
\subsection{L19}
\begin{itemize}
\item 4)
 In the case of $L19$,  define $\tilde j_4=j_0+s_4, x_4=\tilde j_1 e_2e_3+\tilde j_2 e_1e_3+\tilde j_3 e_1e_2+\tilde j_4e_2e_4$, $\bar x_3=\tilde j_1 \overline{e_2e_3}+\tilde j_2\overline{e_1e_3}+\tilde j_3\overline{e_1e_2} +\tilde j_4\overline{e_2e_4}$.

Define $X_4=\left(\begin{array}{cc} 
 x_4 ,& x_4\bar x_4\\
I_4& \bar x_4\end{array}\right)$,  $V_4=\left(\begin{array}{cc}
 a_1 I_4-a_2 e_1e_2& 0\\
 0&d_2 e_1e_2\end{array}\right)$ and\\ $V^\dagger_4=\left(\begin{array}{cc}
-d_2\overline{e_1e_2}&0\\
0&a_1\bar I_4-a_2\overline{e_1e_2}\end{array}\right)$

Calculate $V_4 X_4 V^\dagger_4-X_4$.

\item 5) Define $\tilde j_5=j_0+s_3, x_5=\tilde j_1 e_2e_3+\tilde j_2 e_1e_3+(\tilde j_3-\tilde j_5) e_1e_2+\tilde j_4e_2e_4$, $\bar x_5=\tilde j_1 \overline{e_2e_3}+\tilde j_2\overline{e_1e_3}+(\tilde j_3-\tilde j_5)\overline{e_1e_2} +\tilde j_4\overline{e_2e_4}$.

Define $X_5=\left(\begin{array}{cc} 
 x_5 ,& x_5\bar x_5\\
I_4& \bar x_5\end{array}\right)$, $V_5=\left(\begin{array}{cc}
 a_1 I_4& 0\\
 0&-d_1 e_2e_4\end{array}\right)$ and $V^\dagger_5=\left(\begin{array}{cc}
-d_1\overline{e_2e_4}&0\\
0&a_1\bar I_4\end{array}\right)$

Calculate $V_5 X_5 V^\dagger_5-X_5$.

\item 6) Define $\tilde j_6=j_0+s_4, x_6=\tilde j_1 e_2e_3+\tilde j_2 e_1e_3+(\tilde j_3-\tilde j_5) e_1e_2+\tilde j_4e_2e_4$, $\bar x_6=\tilde j_1 \overline{e_2e_3}+\tilde j_2\overline{e_1e_3}+(\tilde j_3-\tilde j_5)\overline{e_1e_2} +(\tilde j_4-\tilde j_6)\overline{e_2e_4}$.

Define $X_6=\left(\begin{array}{cc} 
 x_6 ,& x_6\bar x_6\\
I_4& \bar x_6\end{array}\right)$,  $V_6=\left(\begin{array}{cc}
 a_1 I_4& -b_1e_2e_3\\
 -c_1 e_2e_3&0\end{array}\right)$ and $V^\dagger_6=\left(\begin{array}{cc}
0&-c_1\overline{e_2e_3}\\
-b_1\overline{e_2e_3}&a_1\bar I_4\end{array}\right)$.

Calculate $V_6 X_6 V^\dagger_6-X_6$.

\item 7) Define $\tilde j_7=j_0+s_1, x_7=(\tilde j_1 -\tilde j_7) e_2e_3+\tilde j_2 e_1e_3+(\tilde j_3-\tilde j_5) e_1e_2+\tilde j_4e_2e_4$, $\bar x_7=(\tilde j_1-\tilde j_7) \overline{e_2e_3}+\tilde j_2\overline{e_1e_3}+(\tilde j_3-\tilde j_5)\overline{e_1e_2} +(\tilde j_4-\tilde j_6)\overline{e_2e_4}$.

Define $X_7=\left(\begin{array}{cc} 
 x_7 ,& x_7\bar x_7\\
I_4& \bar x_7\end{array}\right)$,  $V_7=\left(\begin{array}{cc}
 a_1 I_4& -b_2e_1e_3\\
 -c_2 e_1e_3&0\end{array}\right)$ and \\$V^\dagger_7=\left(\begin{array}{cc}
0&-c_2\overline{e_1e_3}\\
-b_2\overline{e_1e_3}&a_1\bar I_4\end{array}\right)$

Calculate $V_7 X_7 V^\dagger_7-X_7$.

\item 8) Define $\tilde j_8=j_0+s_2, x_8=(\tilde j_1 -\tilde j_7) e_2e_3+(\tilde j_2-\tilde j_8) e_1e_3+(\tilde j_3-\tilde j_5) e_1e_2+\tilde j_4e_2e_4$, $\bar x_8=(\tilde j_1-\tilde j_7) \overline{e_2e_3}+(\tilde j_2-\tilde j_8)\overline{e_1e_3}+(\tilde j_3-\tilde j_5)\overline{e_1e_2} +(\tilde j_4-\tilde j_6)\overline{e_2e_4}$. 

Define $X_8=\left(\begin{array}{cc} 
 x_8 ,& x_8\bar x_8\\
I_4& \bar x_8\end{array}\right)$, $V_8=\left(\begin{array}{cc}
 a_1 I_4& -b_1e_2e_3\\
 -c_1 e_2e_3&0\end{array}\right)$ and $V^\dagger_8=\left(\begin{array}{cc}
0&-c_1\overline{e_2e_3}\\
-b_1\overline{e_2e_3}&a_1\bar I_4\end{array}\right)$.

Calculate $V_8 X_8 V^\dagger_8-X_8$.

\item 9) In the case of step 3 and step 4, the variance of average of absolute value of eigenvalues are large. The large variance can be reduced by the singular value decomposition (SVD) method\cite{LN07}.
\end{itemize}

\subsection{L20}
\begin{itemize}
\item 4-5) Same as $L19$

\item 6) Define $\tilde j_6=j_0+s_2, x_6=(\tilde j_1-\tilde j_5) e_2e_3+(\tilde j_2-\tilde j_6) e_1e_3+\tilde j_3 e_1e_2+\tilde j_4e_2e_4$, $\bar x_6=(\tilde j_1-\tilde j_5) \overline{e_2e_3}+(\tilde j_2-\tilde j_6)\overline{e_1e_3}+\tilde j_3\overline{e_1e_2} +\tilde j_4\overline{e_2e_4}$.

Define $X_6=\left(\begin{array}{cc} 
 x_6 ,& x_6\bar x_6\\
I_4& \bar x_6\end{array}\right)$,  $V_6=\left(\begin{array}{cc}
 a_1 I_4& -b_2 e_1e_3\\
 -c_2e_1e_3&0\end{array}\right)$ and $V^\dagger_6=\left(\begin{array}{cc}
0&-c_2\overline{e_1e_3}\\
-b_2\overline{e_1e_3}&a_1\bar I_4\end{array}\right)$

Calculate $V_6 X_6 V^\dagger_6-X_6$.

\item 7) Define $\tilde j_7=j_0+s_3, x_7=(\tilde j_1-\tilde j_5) e_2e_3+(\tilde j_2-\tilde j_6) e_1e_3+(\tilde j_3-\tilde j_7) e_1e_2+\tilde j_4e_2e_4$, $\bar x_7=(\tilde j_1-\tilde j_5) \overline{e_2e_3}+(\tilde j_2-\tilde j_6)\overline{e_1e_3}+(\tilde j_3 -\tilde j_7)\overline{e_1e_2} +\tilde j_4\overline{e_2e_4}$.

Define $X_7=\left(\begin{array}{cc} 
 x_7 ,& x_7\bar x_7\\
I_4& \bar x_7\end{array}\right)$,  $V_7=\left(\begin{array}{cc}
 a_1 I_4-a_2 e_1e_2&0\\
0& -d_2e_1e_2\end{array}\right)$ and \\$V^\dagger_7=\left(\begin{array}{cc}
-d_2\overline{e_1e_2}&0\\
0&a_1\bar I_4-a_2\overline{e_1e_2}\end{array}\right)$

Calculate $V_7 X_7 V^\dagger_7-X_7$.

\item 8) Define $\tilde j_8=j_0+s_4, x_8=(\tilde j_1-\tilde j_5) e_2e_3+(\tilde j_2-\tilde j_6) e_1e_3+(\tilde j_3-\tilde j_7) e_1e_2+\tilde j_4e_2e_4$, $\bar x_8=(\tilde j_1-\tilde j_5) \overline{e_2e_3}+(\tilde j_2-\tilde j_6)\overline{e_1e_3}+(\tilde j_3 -\tilde j_7)\overline{e_1e_2} +(\tilde j_4-\tilde j_8)\overline{e_2e_4}$.

Define $X_8=\left(\begin{array}{cc} 
 x_8 ,& x_8\bar x_8\\
I_4& \bar x_8\end{array}\right)$,  $V_8=\left(\begin{array}{cc}
 a_1 I_4&0\\
0& -d_1e_2e_4\end{array}\right)$ and $V^\dagger_8=\left(\begin{array}{cc}
-d_1\overline{e_2e_4}&0\\
0&a_1\bar I_4\end{array}\right)$

Calculate $V_8 X_8 V^\dagger_8-X_8$.
\end{itemize}

\begin{figure*}
\begin{minipage}{0.47\linewidth}
\begin{center}
\includegraphics[width=8cm,angle=0,clip]{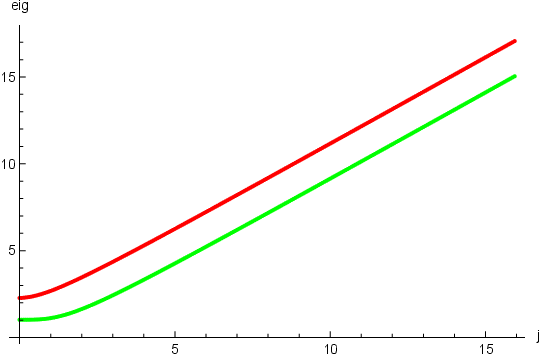}
\end{center}
\end{minipage}
\quad
\begin{minipage}{0.47\linewidth}
\begin{center}
\includegraphics[width=8cm,angle=0,clip]{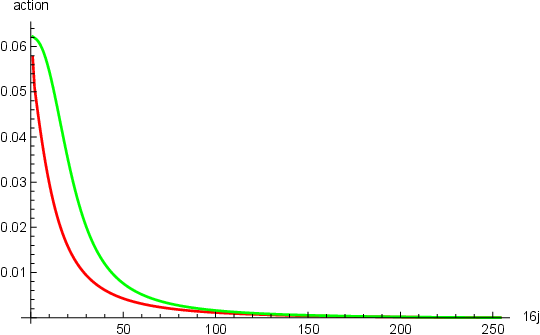}
\end{center}
\end{minipage}

\caption{Eigenvalues (left) and actions (right) of $L19,$ $L20,$ $L25$ and $L21,L22$ with $e_2 e_4$ step 3 after Singular Value Decomposition (SVD). Red points are contribution of large eigenvalues, green points are contribution of small eigenvalues.
The value $j$ is the coordinate of $u_1$. $u_2=3$. }
\end{figure*}

\begin{figure*}
\begin{minipage}{0.47\linewidth}
\begin{center}
\includegraphics[width=8cm,angle=0,clip]{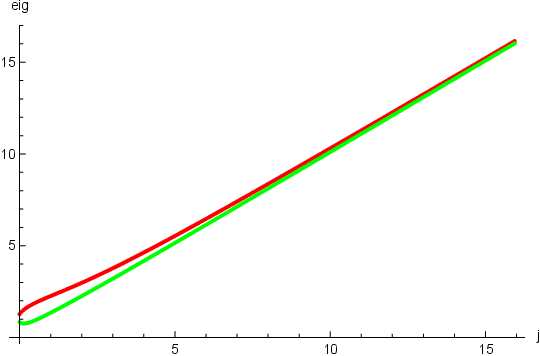}
\end{center}
\end{minipage}
\quad
\begin{minipage}{0.47\linewidth}
\begin{center}
\includegraphics[width=8cm,angle=0,clip]{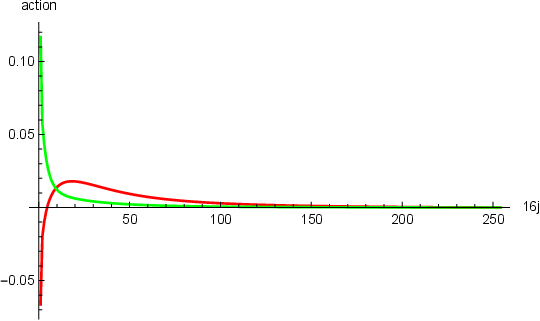}
\end{center}
\end{minipage}

\caption{Eigenvalues (left) and actions (right) of $L23,L24$ and $L21,L22$ with $e_1e_4$ step 3 after SVD. Red lines are the contribution of large eigenvalues, green lines are contribution of small eigenvalues.
The value $j$ is the coordinate of $u_1$. $u_2=1$}
\end{figure*}
\begin{figure*}[htb]
\begin{minipage}{0.47\linewidth}
\begin{center}
\includegraphics[width=6cm,angle=0,clip]{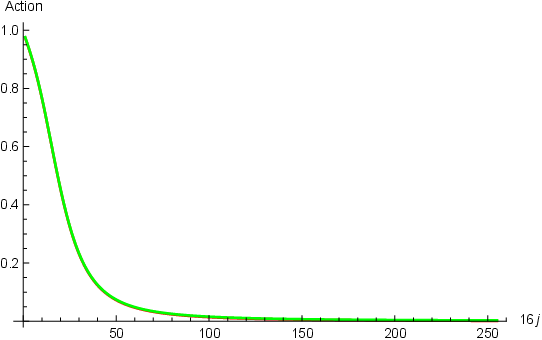}
\end{center}
\end{minipage}
\quad
\begin{minipage}{0.47\linewidth}
\begin{center}
\includegraphics[width=6cm,angle=0,clip]{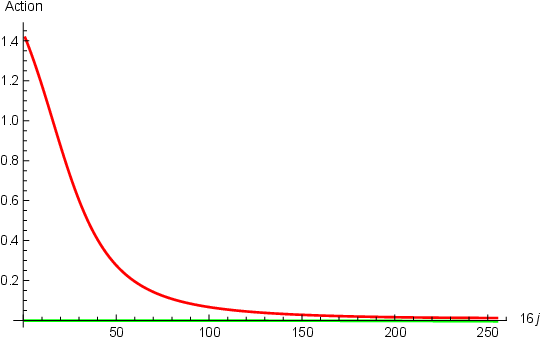}
\end{center}
\label{hysteresis}
\end{minipage}
\caption{The action of $L22$ step 12 (left) and $L24$ step 10 as a function of $16j=0,\cdots, 255$. Red lines are the contribution of large eigenvalues, green lines are the contribution of small eigenvalues. $u_2=3$}
\end{figure*}

\subsection{L21}
The step 6 is $-e_3e_4$. The step 3 is $e_1e_4$ or $e_2e_4$, but mixing of different $e_ke_4$ causes mixing of real coordinate and imaginary coordinate. Therefore we double the number of steps and consider $e_1e_4,-e_1e_4$ and $e_2e_4,-e_2e_4$ pairs as shown in Table 2.. 

\begin{itemize}
\item 4) Define $\tilde j_4=j_0+s_4, x_4=\tilde j_1e_2e_3+\tilde j_2 e_1e_3+\tilde j_3e_1e_2+\tilde j_4 e_3e_4$, $\bar x_4=\tilde j_1\overline{e_2e_3}+\tilde j_2\overline{ e_1e_3}+\tilde j_3\overline{e_1e_2}+\tilde j_4 \overline{e_3e_4}$,

Define $X_4=\left(\begin{array}{cc} 
 x_4 ,& x_4\bar x_4\\
I_4& \bar x_4\end{array}\right)$,  $V_4=\left(\begin{array}{cc}
 a_1 I_4&0\\
0& d_1e_3e_4\end{array}\right)$ and $V^\dagger_4=\left(\begin{array}{cc}
d_1\overline{e_3e_4}&0\\
0&a_1\bar I_4\end{array}\right)$

Calculate $V_4 X_4 V^\dagger_4-X_4$.
\item 5) Same as $L19$.
\item 6) Same as $L19$.

\item 7) Define $\tilde j_7=j_0+s_4$, $x_7=(\tilde j_1-\tilde j_6)e_2e_3+\tilde j_2 e_1e_3+(\tilde j_3-\tilde j_5)e_1e_2+(\tilde j_4 -\tilde j_7)e_3e_4$, $\bar x_4=(\tilde j_1-\tilde j_6)\overline{e_2e_3}+\tilde j_2\overline{ e_1e_3}+(\tilde j_3-\tilde j_5)\overline{e_1e_2}+(\tilde j_4 -\tilde j_7)\overline{e_3e_4}$,

Define $X_7=\left(\begin{array}{cc} 
 x_7 ,& x_7\bar x_7\\
I_4& \bar x_7\end{array}\right)$,  $V_7=\left(\begin{array}{cc}
 a_1 I_4&0\\
0&-d_1e_3e_4\end{array}\right)$ and $V^\dagger_7=\left(\begin{array}{cc}
-d_1\overline{e_3e_4}&0\\
0&a_1\bar I_4\end{array}\right)$

Calculate $V_7 X_7 V^\dagger_7-X_7$.

\item 8) Define $\tilde j_8=j_0+s_2$, $x_8=(\tilde j_1-\tilde j_6)e_2e_3+(\tilde j_2-\tilde j_8) e_1e_3+(\tilde j_3-\tilde j_5) e_1e_2+(\tilde j_4 -\tilde j_7)e_3e_4$, $\bar x_4=(\tilde j_1-\tilde j_6)\overline{e_2e_3}+(\tilde j_2-\tilde j_8)\overline{ e_1e_3}+(\tilde j_3-\tilde j_5)\overline{e_1e_2}+(\tilde j_4 -\tilde j_7)\overline{e_3e_4}$,

Define $X_8=\left(\begin{array}{cc} 
 x_8 ,& x_8\bar x_8\\
I_4& \bar x_8\end{array}\right)$,  $V_8=\left(\begin{array}{cc}
 a_1 I_4&-b_2 e_1e_3\\
-c_2e_1e_3&0\end{array}\right)$ and $V^\dagger_8=\left(\begin{array}{cc}
0&-c_2\overline{e_1e_3}\\
-b_2\overline{e_1e_3}&a_1\bar I_4\end{array}\right)$

Calculate $V_8 X_8 V^\dagger_8-X_8$.
\end{itemize}

\subsection{L22}
\begin{itemize}
\item 4) 
In the step 3, there are two possibilities $e_1e_4$ or $e_2e_4$. The latter is same as $L19$ and the former is same as $L21$. 

\item 5)
The step 4 is same as $L19$ and $L21$

\item 6) Define $\tilde j_6=j_0+s_1$, 

$x_6=(\tilde j_1-\tilde j_6) e_2e_3+\tilde j_2 e_1e_3+(\tilde j_3-\tilde j_5) e_1e_2+\tilde j_4e_2e_4$, $\bar x_6=(\tilde j_1 -\tilde j_6)\overline{e_2e_3}+\tilde j_2\overline{e_1e_3}+(\tilde j_3-\tilde j_5)\overline{e_1e_2} +\tilde j_4\overline{e_2e_4}$.

$x_6=(\tilde j_1-\tilde j_6) e_2e_3+\tilde j_2 e_1e_3+(\tilde j_3-\tilde j_5) e_1e_2+\tilde j_4e_1e_4$, $\bar x_5=(\tilde j_1 -\tilde j_6)\overline{e_2e_3}+\tilde j_2\overline{e_1e_3}+(\tilde j_3-\tilde j_5)\overline{e_1e_2} +\tilde j_4\overline{e_1e_4}$

Define $X_6=\left(\begin{array}{cc} 
 x_6 ,& x_6\bar x_6\\
I_4& \bar x_6\end{array}\right)$, $V_6=\left(\begin{array}{cc}
 a_1 I_4&-b_1 e_2e_3\\
 -c_1 e_2e_3&0\end{array}\right)$ and $V^\dagger_6=\left(\begin{array}{cc}
0&-c_1\overline{e_2e_3}\\
-b_1\overline{e_2e_3}&a_1\bar I_4\end{array}\right)$.

Calculate $V_6 X_6 V^\dagger_6-X_6$.
\item 7)
Define $\tilde j_7=j_0+s_2$, 

$x_7=(\tilde j_1-\tilde j_6) e_2e_3+(\tilde j_2-\tilde j_7) e_1e_3+(\tilde j_3-\tilde j_5) e_1e_2+\tilde j_4e_2e_4$, $\bar x_7=(\tilde j_1 -\tilde j_6)\overline{e_2e_3}+\tilde j_2\overline{e_1e_3}+(\tilde j_3-\tilde j_5)\overline{e_1e_2} +\tilde j_4\overline{e_2e_4}$.

$x_7=(\tilde j_1-\tilde j_6) e_2e_3+(\tilde j_2-\tilde j_7) e_1e_3+(\tilde j_3-\tilde j_5) e_1e_2+\tilde j_4e_1e_4$, $\bar x_7=(\tilde j_1 -\tilde j_6)\overline{e_2e_3}+(\tilde j_2-\tilde j_7)\overline{e_1e_3}+(\tilde j_3-\tilde j_5)\overline{e_1e_2} +\tilde j_4\overline{e_1e_4}$

Define $X_7=\left(\begin{array}{cc} 
 x_7 ,& x_7\bar x_7\\
I_4& \bar x_7\end{array}\right)$, $V_7=\left(\begin{array}{cc}
 a_1 I_4&-b_2 e_1e_3\\
 -c_2 e_1e_3&0\end{array}\right)$ and $V^\dagger_7=\left(\begin{array}{cc}
0&-c_2\overline{e_1e_3}\\
-b_2\overline{e_1e_3}&a_1\bar I_4\end{array}\right)$

Calculate $V_7 X_7 V^\dagger_7-X_7$.
\item 8) Define $j_8=j_0+s_4$.
$x_8=(\tilde j_1-\tilde j_6) e_2e_3+(\tilde j_2-\tilde j_7) e_1e_3+(\tilde j_3-\tilde j_5) e_1e_2+\tilde j_4e_2e_4-\tilde j_8 e_3e_4$, $\bar x_8=(\tilde j_1 -\tilde j_6)\overline{e_2e_3}+\tilde j_2\overline{e_1e_3}+(\tilde j_3-\tilde j_5)\overline{e_1e_2} +\tilde j_4\overline{e_2e_4}-\tilde j_8\overline{e_3e_4}$.

Define $X_8=\left(\begin{array}{cc} 
 x_8 ,& x_8\bar x_8\\
I_4& \bar x_8\end{array}\right)$, $V_8=\left(\begin{array}{cc}
 a_1 I_4&0\\
0& -d_1 e_3e_4\end{array}\right)$ and $V^\dagger_8=\left(\begin{array}{cc}
d_1\overline{e_3e_4}&0\\
0&a_1\bar I_4\end{array}\right)$

Calculate $V_8 X_8 V^\dagger_8-X_8$.
\end{itemize}

\subsection{L23}
\begin{itemize}
\item 4)
The step 3 is $e_1e_4$ but the step 6 is $-e_2e_4$.  We double the number of steps as shown in the Table 2. 
\item 5) Define $\tilde j_5=j_0+s_2$, 

$x_5=\tilde j_1 e_2e_3+(\tilde j_2 -\tilde j_5)e_1e_3+\tilde j_3 e_1e_2+\tilde j_4e_2e_4$, $\bar x_5=\tilde j_1\overline{e_2e_3}+(\tilde j_2-\tilde j_5)\overline{e_1e_3}+\tilde j_3\overline{e_1e_2} +\tilde j_4\overline{e_3e_4}$.
\item 6)
Define $\tilde j_6=j_0+s_1$,
$x_6=(\tilde j_1-\tilde j_6) e_2e_3+(\tilde j_2-\tilde j_5) e_1e_3+\tilde j_3 e_1e_2+\tilde j_4e_1e_4$, $\bar x_5=(\tilde j_1 -\tilde j_6)\overline{e_2e_3}+\tilde j_2\overline{e_1e_3}+(\tilde j_3-\tilde j_5)\overline{e_1e_2} +\tilde j_4\overline{e_3e_4}$

Define $X_6=\left(\begin{array}{cc} 
 x_6 ,& x_6\bar x_6\\
I_4& \bar x_6\end{array}\right)$, $V_6=\left(\begin{array}{cc}
 a_1 I_4&-b_2 e_1e_3\\
 -c_2 e_1e_3&0\end{array}\right)$ and $V^\dagger_6=\left(\begin{array}{cc}
0&-c_2\overline{e_1e_3}\\
-b_2\overline{e_1e_3}&a_1\bar I_4\end{array}\right)$.

Calculate $V_6 X_6 V^\dagger_6-X_6$.

\item 7) Define $\tilde j_7=j_0+s_4$,
$x_7=(\tilde j_1-\tilde j_6) e_2e_3+(\tilde j_2 -\tilde j_5)e_1e_3+\tilde j_3 e_1e_2+\tilde j_4e_1e_4$, $\bar x_7=(\tilde j_1 -\tilde j_6)\overline{e_2e_3}+\tilde j_2\overline{e_1e_3}+\tilde j_3\overline{e_1e_2} +(\tilde j_4-\tilde j_7)\overline{e_3e_4}$

Define $X_7=\left(\begin{array}{cc} 
 x_7 ,& x_7\bar x_7\\
I_4& \bar x_7\end{array}\right)$, $V_7=\left(\begin{array}{cc}
 a_1 I_4&0\\
 0&-d_1 e_3e_4\end{array}\right)$ and $V^\dagger_7=\left(\begin{array}{cc}
-d_1\overline{e_3e_4}&0\\
0&a_1\bar I_4\end{array}\right)$.

Calculate $V_7 X_7 V^\dagger_7-X_7$.
\item 8) Define $\tilde j_8=j_0+s_3$,
$x_8=(\tilde j_1-\tilde j_6) e_2e_3+(\tilde j_2 -\tilde j_5)e_1e_3+(\tilde j_3-\tilde j_8) e_1e_2+(\tilde j_4-\tilde j_7)e_3e_4$, $\bar x_8=(\tilde j_1 -\tilde j_6)\overline{e_2e_3}+\tilde j_2\overline{e_1e_3}+(\tilde j_3-\tilde j_8)\overline{e_1e_2} +(\tilde j_4-\tilde j_7)\overline{e_3e_4}$.

 Define $X_8=\left(\begin{array}{cc} 
 x_8 ,& x_8\bar x_8\\
I_4& \bar x_8\end{array}\right)$, $V_8=\left(\begin{array}{cc}
 a_1 I_4-a_2 e_1e_2&0\\
0& -d_2 e_1e_2\end{array}\right)$ and \\$V^\dagger_8=\left(\begin{array}{cc}
-d_2\overline{e_1e_2}&0\\
0&a_1\bar I_4-a_2 \overline{e_1e_2}\end{array}\right)$.

Calculate $V_8 X_8 V^\dagger_8-X_8$.
\end{itemize}

\subsection{L24}
Up to the step 5 $L24=L23$.
\begin{itemize}
\item 7) Define $\tilde j_7=j_0+s_3$,
$x_7=(\tilde j_1-\tilde j_6) e_2e_3+(\tilde j_2 -\tilde j_5)e_1e_3+(\tilde j_3-\tilde j_7) e_1e_2+\tilde j_4e_3e_4$, $\bar x_7=(\tilde j_1 -\tilde j_6)\overline{e_2e_3}+\tilde j_2\overline{e_1e_3}+(\tilde j_3-\tilde j_7)\overline{e_1e_2} +\tilde j_4\overline{e_3e_4}$

Define $X_7=\left(\begin{array}{cc} 
 x_7 ,& x_7\bar x_7\\
I_4& \bar x_7\end{array}\right)$, $V_7=\left(\begin{array}{cc}
 a_1 I_4-a_2 e_1e_2&0\\
 0&-d_2 e_1e_2\end{array}\right)$ and \\$V^\dagger_7=\left(\begin{array}{cc}
-d_2\overline{e_1e_2}&0\\
0&a_1\bar I_4-a_2\overline{e_1e_2}\end{array}\right)$.

Calculate $V_7 X_7 V^\dagger_7-X_7$.
\item 8) Define $\tilde j_8=j_0+s_4$,
$x_8=(\tilde j_1-\tilde j_6) e_2e_3+(\tilde j_2 -\tilde j_5)e_1e_3+(\tilde j_3-\tilde j_7) e_1e_2+(\tilde j_4-\tilde j_8)e_3e_4$, $\bar x_8=(\tilde j_1 -\tilde j_6)\overline{e_2e_3}+\tilde j_2\overline{e_1e_3}+(\tilde j_3-\tilde j_7)\overline{e_1e_2} +(\tilde j_4-\tilde j_8)\overline{e_3e_4}$

Define $X_8=\left(\begin{array}{cc} 
 x_8 ,& x_8\bar x_8\\
I_4& \bar x_8\end{array}\right)$, $V_8=\left(\begin{array}{cc}
 a_1 I_4&0\\
 0&-d_1 e_3e_4\end{array}\right)$ and $V^\dagger_8=\left(\begin{array}{cc}
-d_1\overline{e_3e_4}&0\\
0&a_1\bar I_4\end{array}\right)$.

Calculate $V_8 X_8 V^\dagger_8-X_8$.

\end{itemize}

\subsection{L25}
\begin{itemize}
\item 4) The step 3 of $L25$ is along $e_2e_4$ and the step 6 is along $-e_2e_4$. 
The transformations are same as $L19$. 
\item 5) Define $\tilde j_5=j_0+s_1, x_5=(\tilde j_1-\tilde j_5) e_2e_3+\tilde j_2 e_1e_3+\tilde j_3 e_1e_2+\tilde j_4e_2e_4$, $\bar x_5=(\tilde j_1-\tilde j_5) \overline{e_2e_3}+\tilde j_2\overline{e_1e_3}+\tilde j_3\overline{e_1e_2} +\tilde j_4\overline{e_2e_4}$.

Define $X_5=\left(\begin{array}{cc} 
 x_5 ,& x_5\bar x_5\\
I_4& \bar x_5\end{array}\right)$, $V_5=\left(\begin{array}{cc}
 a_1 I_4& -b_1 e_2e_3\\
 -c_1e_2e_3&0\end{array}\right)$ and \\$V^\dagger_5=\left(\begin{array}{cc}
0&-c_1\overline{e_2e_3}\\
-b_1\overline{e_2e_3}&a_1\bar I_4\end{array}\right)$

Calculate $V_5 X_5 V^\dagger_5-X_5$.
\item 6) Define $\tilde j_6=j_0+s_2, x_6=(\tilde j_1-\tilde j_5) e_2e_3+(\tilde j_2-\tilde j_6) e_1e_3+\tilde j_3 e_1e_2+\tilde j_4e_2e_4$, $\bar x_6=(\tilde j_1-\tilde j_5) \overline{e_2e_3}+(\tilde j_2-\tilde j_6)\overline{e_1e_3}+\tilde j_3\overline{e_1e_2} +\tilde j_4\overline{e_2e_4}$.

Define $X_6=\left(\begin{array}{cc} 
 x_6 ,& x_6\bar x_6\\
I_4& \bar x_6\end{array}\right)$, $V_6=\left(\begin{array}{cc}
 a_1 I_4& -b_2 e_1e_3\\
 -c_2e_1e_3&0\end{array}\right)$ and $V^\dagger_6=\left(\begin{array}{cc}
0&-c_2\overline{e_1e_3}\\
-b_2\overline{e_1e_3}&a_1\bar I_4\end{array}\right)$

Calculate $V_6 X_6 V^\dagger_6-X_6$.

\item 7) Define $\tilde j_7=j_0+s_3, x_7=(\tilde j_1-\tilde j_5) e_2e_3+(\tilde j_2-\tilde j_6) e_1e_3+(\tilde j_3-\tilde j_7) e_1e_2+\tilde j_4e_2e_4$, $\bar x_7=(\tilde j_1-\tilde j_5) \overline{e_2e_3}+(\tilde j_2-\tilde j_6)\overline{e_1e_3}+\tilde j_3\overline{e_1e_2} +\tilde j_4\overline{e_2e_4}$.

Define $X_7=\left(\begin{array}{cc} 
 x_7 ,& x_7\bar x_7\\
I_4& \bar x_7\end{array}\right)$, $V_7=\left(\begin{array}{cc}
 a_1 I_4-a_2 e_1e_2& 0\\
0&-d_2e_1e_2\end{array}\right)$ and \\$V^\dagger_7=\left(\begin{array}{cc}
-d_2\overline{e_1e_2}&0\\
0&a_1\bar I_4-a_2\overline{e_1e_2}\end{array}\right)$

Calculate $V_7 X_7 V^\dagger_7-X_7$.
 
\item 8) Define $\tilde j_8=j_0+s_4, x_8=(\tilde j_1-\tilde j_5) e_2e_3+(\tilde j_2-\tilde j_6) e_1e_3+(\tilde j_3-\tilde j_7) e_1e_2+(\tilde j_4-\tilde j_8)e_2e_4$, $\bar x_8=(\tilde j_1-\tilde j_5) \overline{e_2e_3}+(\tilde j_2-\tilde j_6)\overline{e_1e_3}+\tilde j_3\overline{e_1e_2} +(\tilde j_4-\tilde j_8)\overline{e_2e_4}$.

Define $X_8=\left(\begin{array}{cc} 
 x_8 ,& x_8\bar x_8\\
I_4& \bar x_8\end{array}\right)$, $V_8=\left(\begin{array}{cc}
 a_1 I_4& 0\\
0&-d_1e_2e_4\end{array}\right)$ and $V^\dagger_8=\left(\begin{array}{cc}
-d_1\overline{e_2e_4}&0\\
0&a_1\bar I_4\end{array}\right)$
Calculate $V_8 X_8 V^\dagger_8-X_8$.
\end{itemize}

\subsection{Numerical  Results}
I calculated the average of absolute value of eigenvalues of $X\bar X$ for time transformation proportional to $e_2e_4$ $(L19,L20,L25)$ and to $e_1e_4$ $(L21,L22,L23,L24)$ at the step 3. I checked that the absolute values of the eigenvalue of $x\bar x$ in ${\mathcal X}$ have a peak in the middle of steps and vanish at the step 8 and the eigenvalues of the $(L19,L20,L25)$ agree, and those of $(L21,L22,L23,L24)$ agree.

The method of SVD was applied in the tensor renormalization \cite{XCQZYX12, AOT20, Akiyama22, YNO23}. A package for Grassmann tensor network written in Python \cite{Yosprakob23} is used in a spacial 2D, flavor 3D QCD simulation\cite{YNO23},
and the simplicity of their algorithm as compared to \cite{XCQZYX12,AOT20,Akiyama22} was remarked. But I guess that the simplicity is due to the fact that the spacial dimension is two. 

One quaternion in $(3+1)D$ simulation is represented by a $4\times 4$ matrix, and there are 4 eigenvalues. In some loops like $L21$, they are nearly equal but in some loops like $L19,L20,L24$ there are 2 kinds of eigenvalues and we need to separate the two groups, in order to reduce the variance.

\subsection{Hysteresis effects}
Mathematica is a powerfull tool to study quaternions. In \cite{GAS06} a formula
\begin{equation}
exp A[{\bf x}]=I+\frac{\sin \ell}{\ell}A[{\bf x}]+\frac{1-\cos\ell}{\ell^2} A[{\bf x}]^2
\end{equation}
where $A[{\bf x}]=\left(\begin{array}{ccc}
0&-z&y\\
z&0&-x\\
-y&x&0\end{array}\right)$ satisfies $A[{\bf x}]{\bf w}={\bf x\times w}$ for all ${\bf w}\in R^3$ is derived using quaternions.

In $(2+1)D$, quaternions ${\bf q}_1=a_1+b_1{\bf i}$, ${\bf q}_2=a_2+b_2{\bf j}$ satisfy ${\bf q}_1{\bf q}_2=a_1a_2+b_1a_2{\bf i}+a_1b_2{\bf j}+b_1b_2{\bf k}$. 

 The matrix of the rotation around $\bf x$ with a counterclockwise angle $|{\bf x}|=\ell$ equals 
\begin{equation}
R[{\bf x}]=\left(\begin{array}{ccc}
1&0&0\\
0&1&f_{23}(\ell)\\
0&f_{32}(\ell)&1\end{array}\right)
\end{equation} 

The $f_{23}(\ell)$ and the $f_{32}(\ell)$ are shown in Fig.\ref{hysteresis}. Evaluation of the action along the path is not easy, but return to the original point can be seen by the fact that $f_{23}(\ell)$ and $f_{32}(\ell)$ forms a closed curve. The time reversal symmetry
in the TR-NEWS will allow derivation of the weight function of $L19-L25$ from actions obtained after the SVD.

\begin{figure*}[htb]
\begin{minipage}{0.47\linewidth}
\begin{center}
\includegraphics[width=6cm,angle=0,clip]{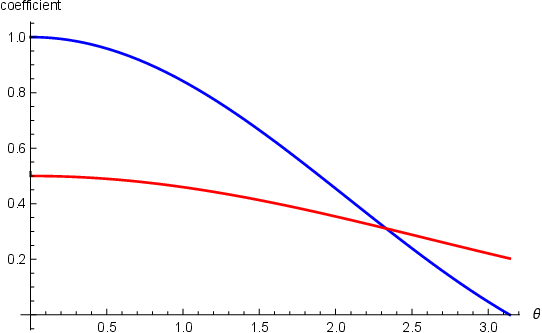}
\end{center}
\caption{The first coefficient (blue) and the second coefficient (red) of $exp A[{\bf x}]$ as functions of $\theta=|{\bf x}|$.}
\end{minipage}
\quad
\begin{minipage}{0.47\linewidth}
\begin{center}
\includegraphics[width=6cm,angle=0,clip]{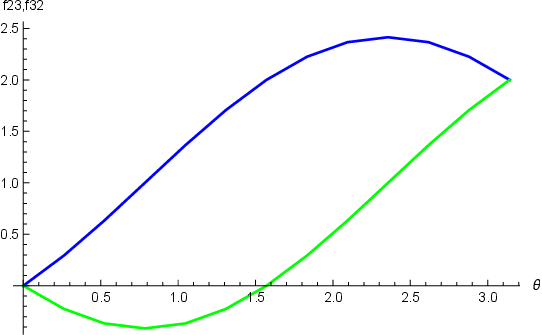}
\end{center}
\caption{$f_{23}$ (green) and $f_{32}$ (blue) as a function of $\ell=\theta$.}
\label{hysteresis}
\end{minipage}
\end{figure*}

 Identifying $\theta$ as the time in a frame, the diagram shows hysteresis effects.

 Concerning the action of $L19-L25$, typical data of $L22$ and $L24$ are shown in Fig. 10. The peak of action of $L24$ step 10, whose
contribution of smaller eigenvalue part is negligible about $\sqrt 2$ times larger than that of $L22$ step 12, whose smaller eigenvalues and the larger eigenvalue are almost degenerate.

In the case of $L19, L20$ and $L25$, actions calculated by the 1st derivative of eigenvalues with respect to the distance from the origin $j$ at steps 6, 7 are 0. In the case of $L21,L22,L23$ and $L24$, actions of steps 5,6,7 and steps 13,14,15 are 0. 

 It is necessary to calculate the convolution of direct wave and the TR wave that propagate from a position of the transducer to the receiver.

\subsection{Rotations in (2+1)D}
We applied the same method of calculing action of B type loops in (2+1)D.

\begin{tabular}{r|cccccccccccc}
 step & 0&1&2&3 &4&5&6&7&8&9&10&11\\
\hline
L3'&x&y&t&-y&-x&-t&x &y&-t&y&x&t\\
&23&31&34/14&-31&-23&-24&-23 &-31&-34/14&31&23&24\\
\hline
L4'&x&y&t&-x&-y&-t& &\\
&23&31&34&-23&-13&-34&&\\
\hline
L7'&x&y&x&t&-y&-x&-x&-t\\
&23&31&23&34&-13&-23&-23&-34\\
\end{tabular}

 Table 3. Directions of the wave front of loops $L3', L4'$ and $L7'$ ($z$ in \cite{DGHHN95} is replaced by $t$)  the (2+1)D space.  In the $e_1,e_2,e_4$ bases (the first line) and in the $e_ie_j$ bases (the second line).
\vskip 0.5 cm
  In these cases, 4 eigenvalues are almost degenerate and the SVD is unnecessary. The link ortogonal to $e_1\wedge e_2$ is chosen to be $e_1e_2$.  
 
When $e_1e_4$ is changed to $e_3e_4$ in $L3'$, the eigenvalues become similar to those of $L21$. After shifting the time axis, the sum of large eigenvalue and the small eigenvalue at a fixed $j$ becomes 0. Hence action becomes 0.

Since the step 2 of $L4'$ and the step 3 of $L7'$ are restricted to $e_3e_4$, the derivative of eigenvalues as a function of $j$ is almost constant and there are no action. Absolute value of eigenvalues of $L'7$ is about twice of $L3'$, since $L7'$ contains the path along $x$ axis twice as long as that of $L3'$.

\section{Perspective of quaternion quantum field theory}
Starting from the extension of the quadratic phase Fourier transform of complex number space of Castro et al.\cite{CMT18} to quaternion number space proposed by Hitzer\cite{Hitzer23}, we presented that the product of quaternions have scalar part and wedge-product parts, and the Dirac's delta function in complex number space cannot be extended to that in quaternion number space as pointed out by Adler in 1985\cite{Adler86}.

We extended the mapping of vectors in $R^2$ space to $R^{2,1}$ space\cite{SFDS23a,SFDS23b,SF23c} using Clifford algebra bases $e_1,e_2, e_1\wedge e_2$ done by Porteous\cite{Porteous95}, to mapping vectors in $R^3$ space to $R^{3,1}$ space, using bases $e_2e_3, e_3 e_1, e_1e_2,$$e_1e_4, e_2e_4, e_3 e_4$ and $\overline{e_2e_3},\overline{e_3 e_1}$,$\overline{e_1 e_2}$,$\overline{e_1e_4}$,$\overline{e_2e_4}$,$\overline{e_3e_4}$.

Porteous obtained actions by sandwitching a state vector of ${\mathcal A}_{2,1}$ expressed by quaternions by Vahlen matrices, which are $2\times 2$ complex matrices.
I extended the state vector to that of ${\mathcal A}_{3,1}$ and adopted Clifford bases given by a product of quaternions, which are $4\times 4$ matrices. Proper actions could be obtained by calculating the difference between the transformed vector and the original vector after singular value decompositions.  

The standard Dirac equation is\cite{Garling11}
\begin{equation}
D_{3,1}\psi=-\gamma_0\frac{\partial \psi}{\partial t}+\gamma_1\frac{\partial \psi}{\partial x_1}+\gamma_2\frac{\partial \psi}{\partial x_2}+\gamma_3\frac{\partial \psi}{\partial x_3},
\end{equation}
where
\begin{eqnarray}
&&\gamma_0=\left(\begin{array}{cc}
0&I\\
1&0\end{array}\right),\quad \gamma_1=\left(\begin{array}{cc}
0&-Q\\
Q&0\end{array}\right),\nonumber\\
&& \gamma_2=\left(\begin{array}{cc}
0&-\sqrt{-1}J\\
\sqrt{-1}J&0\end{array}\right),\quad \gamma_3=\left(\begin{array}{cc}
0&-U\\
U&0\end{array}\right)
\end{eqnarray}

In QED, the Lagrangian is
\begin{equation}
{\mathcal L}_{QED}=\sqrt{-1}\bar\psi\gamma_\mu(\partial_\mu \psi(x)-\sqrt{-1}eA_\mu(x)\psi(x))-m\bar\psi(x)\psi(x)
\end{equation}
is invariant under 
\begin{eqnarray}
&&\psi(x)\to e^{+\sqrt{-1}e\Lambda(x)}\psi(x)\\
&&\bar\psi(x)\to e^{-\sqrt{-1}e\Lambda(x)}\bar\psi(x)\\
&&A^\mu(x)\to A^\mu(x)+\partial^\mu\Lambda(x).
\end{eqnarray}

Replacement of the Dirac $\gamma(e_i)$ to $\gamma(e_ie_j)$ and to choose proper gauge transformations remains a problem.

Another problem related to NDT is incorpolation of the time dependence, or hysteresis effects which can be simulated by the Preisach-Mayergoyz model\cite{PKDS12}. 

Cartan \cite{Cartan66} showed in $\nu=3$ Euclidean space, four sets of 3 vectors can be represented by a linear combination of two products of spinors having quaternion bases, and in $\nu=4$ Euclidean space, two sets of 4 vectors are coupled to a linear sets of four products of spinors having quaternion bases to form invariants.  Adler \cite{Adler86} presented condition for constructing quaternion quantum field theory, and the $\nu=3$ and $\nu=4$ case match the condition. 

We considered 7 FP actions of DeGrand et al\cite{DGHHN95} in $R^{3,1}$ lattice space. The rule of continuation of the direction vector of wave fronts in space-time, we found 12 links containing $\pm e_1e_4$ or $\pm e_2e_4$ and 2 links containing $-e_3e_4$.

\subsection{Quaternion Fourier Transform and Quaternion Domain Fourier Transform}
Hitzer and Sangwine\cite{HS13} expressed a quaternion $q$ as
\begin{eqnarray}
q&=&q_r+q_i {\bf i}+q_j{\bf j}+q_k{\bf k}=q_r+\sqrt{q_i^2+q_j^2+q_k^2}{\bf \mu}(q)\nonumber\\
&=&\cos\alpha+{\bf \mu}(q)\sin\alpha=e^{\alpha{\bf \mu}(q)}
\end{eqnarray}
where ${\bf \mu}(q)=\frac{q_i{\bf i}+q_j{\bf j}+q_k{\bf k}}{\sqrt{q_i^2+q_j^2+q_k^2}}$, $\cos\alpha=q_r$ and $\sin\alpha=\sqrt{q_i^2+q_j^2+q_k^2}$.
Orthogonal 2D planes split is
\begin{equation}
q_\pm=\{q_r\pm q_k+{\bf i}(q_i\mp q_j)\}\frac{1\pm {\bf k}}{2}=\frac{1\pm{\bf k}}{2}\{q_r\pm r_k+{\bf j}(q_j\mp q_i)\}.
\end{equation}

Quaternion Fourier transform of a function $h\in L^1(R^2,{\bf H})$ described by pure quaternions $f,g$ is
\begin{equation}
{\mathcal F}^{f,g}\{h\}(\omega)=\int_{R^2} e^{-f x_1\omega_1}h(x)e^{-g x_2\omega_2} dx_1 dx_2
\end{equation}

The method was extended to quaternion valued quaternion domain Fourier transform (QDFT)\cite{Hitzer16} and applied to an extension of quadratic phase Fourier transform of \cite{CMT18} in the recent article as Quadratic-Phase Quaternion Domain Fourier Transform (QPQDFT)\cite{Hitzer23}.

In the textbook\cite{Hitzer22} the space-time algebra $Cl_{3,1}$ of Minkowski space $R^{3,1}$ with bases $\{e_1,e_2,e_3,e_t\}$ which have the relation
$e_1^2=e_2^2=e_3^2=-e_t^2=1$ isomorphic to a system with bases $\{1, e_t, i_3,i_{st}\}$ where
\begin{equation}
i_3=e_1 e_2 e_3=e_t^*=e_t i_3^{-1}, \quad i_{st}=e_t i_3,\quad i_{st}^2=-1,
\end{equation} 
was introduced.

Hitzer defined the volume-time Fourier transform 
\begin{equation}
{\mathcal F}_{VT}(\omega)=\int_{R^{3,1}} e^{-e_t \omega_t}h({\bf x})e^{-\vec x\cdot \vec\omega} d^4 x
\end{equation}
where
\begin{eqnarray}
{\bf x}&=&t e_t+\vec x,\quad \vec x=x_1 e_1 +x_2 e_2+ x_3 e_3,\nonumber\\
\omega&=&\omega_t e_t+\vec\omega, \quad \vec\omega=\omega_1 e_1+\omega_2 e_2+\omega_3 e_3
\end{eqnarray}

The bases of ${\mathcal A}_{3,1}^+$ taken by Garling\cite{Garling11} are
\begin{equation}
e_1e_3, e_2 e_4, e_1 e_4, e_2e_3, e_3 e_4, e_1e_2,e_\Omega,
\end{equation}
and the bases $e_i e_4$ and $e_j e_4$ ($i\ne j $) do not commute, in contrast to Hitzer's $e_t$ in his volume-time Clifford Fourier transform.

In order to calculate actions in quaternion field theory in $R^{3,1}$, it is necessary to consider the system is isomorphic to $M_2({\bf H})$. The biquaternion space was excluded in Ariel's approach\cite{Ariel21,Ariel23}.

\subsection{Application of (3+1)D quaternion Fourier transform}

The (3+1)D quaternion Fourier transform allow 3D image analysis. In present NDT technology, essentially (2+1)D images are analized, but using two quaternion bases $e_ie_j$ and $\overline{e_i e_j}$, three dimensional detection of the scattering position of ultrasonic waves may become possible. In the fixed point action analysis, we found that variance of actions along $e_ie_4$ becomes large, but via the SVD we can decomose the large eigenvalue component and the small eigenvalue component. Recent NDT experiment using water tanks is presented in \cite{DDSKP23}.

Felsberg and Sommer\cite{FS01} defined in the Riesz wavelet  transform of $(2+1)D$ 
\begin{equation}
G_M({\bf u})=G_3(u_1,u_2,0)-{\bf i}G_1(u_1,u_2,0)-{\bf j}G_2(u_1,u_2,0),
\end{equation}
and applied the Radon transform in $R^3$ by choosing a hyper plane $\xi(\omega,p)=\{ x\in R^3\mid (x,\omega)=p \}$, where
$\omega=(\lambda_1,\lambda_2,\lambda_3)$ is a unit vector, $(x,\omega)=\sum_i x_i\lambda_i$.

For a function $f\in R^3$, $\hat f(\xi)=\hat f(\xi,p)=\int_\xi f(x)d_\xi x$ is called the Fourier transform on Radon measure  or Radon transform\cite{KF65, Horvath66, Iwanami70}.  Here,
$d_\xi x\wedge d\omega=dx$ is the volume element.

We define
\begin{eqnarray}
G_{M 1}({\bf u})&=&G_0({\bf u})-e_2e_3 G_{23}({\bf u})-e_1 e_3 G_{13}({\bf u})-e_1e_2 G_{12}({\bf u})-e_1e_4 G_{14}({\bf u})\nonumber\\
G_{M 2}({\bf u})&=&G_0({\bf u})-e_2e_3 G_{23}({\bf u})-e_1 e_3 G_{13}({\bf u})-e_1e_2 G_{12}({\bf u})-e_2e_4 G_{24}({\bf u})\nonumber\\
G_{M 3}({\bf u})&=&G_0({\bf u})-e_2e_3 G_{23}({\bf u})-e_1 e_3 G_{13}({\bf u})-e_1e_2 G_{12}({\bf u})-e_3e_4 G_{34}({\bf u}).
\end{eqnarray}

By defining the space of hyperplanes as ${\mathcal F}$, the Plancherel's formula says \cite{KF65,Iwanami70}
\begin{equation}
F_M=\int_M \mid f (x)\mid^2 dx=\int_{\mathcal F} \mid\wedge \hat f (\xi)\mid^2 d\xi.
\end{equation}
where $f(x)=\frac{dF}{dx}$ is unique, when $F_M$ is additive. For complex valued function in $L^2$ space the formula is established \cite{Ito65}, but for quaternion valued functions $M_2({\bf H})$, it is not.  

Since numerical results suggest that $F_{M3}\ne F_{M1}=F_{M2}$, the topology of $M_3$ is different from that of $M_1$ and $M_2$.  In (2+1)D QED \cite{ANPRS20}, suggests that one local quaternion flame among 3 frames is unphysical frame containing ghosts. In the case of phonon propagation, the conditions are different\cite{SFDS23}.
In (3+1)D, existence of hysteresis effects suggests presence of local quaternion times in the biquaternion framework.

In mathematics of lattice topology, there remain problems.
A complex torus $T^2=C^2/G$ where $G$ is a discrete subgroup becomes a Hopf manifold $S^1\times S^3$. Choosing $W=C^2-(0,0)$,
$M=W/G=R/Z\times S^3$ is a Hopf manifold\cite{Kodaira92,Hopf32}, and one cannot impose K\"ohrer structure, which is necessary for Hamiltonian dyamics.
In projective space $P^3$, one can identify $z\in W$ and $-z\in W$, one can impose K\"ohrer structure. We start from TR symmetric space and consider appearance of a phase $\delta$ from non-commutativity of $q_1,q_2\in S^3$ as $q_1 q_2=q_2 q_1e^{\sqrt{-1}\delta}$.

The loop depends on the time sequence $(t_1,t_2.\cdots t_m)$, but patching neighborhoods of the local coordinate $(z_j,t)=(z_j^1,z_j^2,t_1,\cdots,t_m)$ defined as the patch $U_j$ is shown to be independent of the time series $t_1,\cdots t_m$ in $P^3$.

 The algebra ${\mathcal A}_{4,1}$ is isomorphic to $M_2({\bf H})\oplus M_2({\bf H})$
\[
j({\mathcal A}_{4,1})=\left(\begin{array}{cc}
x_2{\bf i}+x_3{\bf j}+x_4{\bf k}&-x_1+x_5\\
x_1+x_5&-x_2{\bf i}-x_3{\bf j}-x_4{\bf k}\end{array}\right)
\]
where $x_i$ are real.

In the Light Front Quantum Chromo Dynamics(LFQCD) of Srivastava and Brodsky \cite{SB01,SB02}
$\tau=(t-z/c)/\sqrt{2}$ corresponds to $(x_1-x_5)/\sqrt 2$. For massless particle, propagators are doubly transverse, i.e. with respect to the gauge direction $n_\mu$ and the chilarity direction $k_\mu$.

 The two $M_2({\bf H})$ represent TR symmetric physical fields, and the BRST ghost fields are decoupled.

In the FP lattice simulation, fermion propagation direction and its orthogonal 2D plane can be characterized by quaternions. If a generalization of Castro et al.'s Fourier transform\cite{CMT18} becomes possible, a progress of the quaternion quantum field theory would occur. 

Bosonic wave propagations in Weyl fermion background and in Dirac fermion background are different. The former is suitable for detecting chiral U(1) symmetry and the latter is suitable for detecting charge U(1) symmetry. 

The Clifford algebra bases for  ${\mathcal A}_{3,1}$ were taken from Garling \cite{Garling11}, but there are differences in definition of time. The time axis $e_1e_4, e_2e_4$ and $e_3e_4$ are contained symmetrically. Our model has qualitative differences between the combination of $e_0$ and $e_3e_4$ and $e_0$ and $e_1e_4, e_2e_4$. The paths $L21-L24$ contain different path $e_ie_4$ and $-e_je_4$. We extended
the Vahlen transformation on a (2+1)D Clifford algebra given by Porteous\cite{Porteous95}
to (3+1)D Clifford algebra, and changed the M\"obius transformation matrix element $xx^-$ to
$\sum_{i=1}^3 X_i \bar X_i$. 

There are three quaternion local times in contrast to one local quaternion time postulated by Ariel\cite{Ariel21,Ariel23}. 
When there are hysteresis effects, it might be necessary to consider M\"obius band structure.


{\bf Acknowledgments}

{\small The author thanks Prof. S. Dos Santos for valuable discussions on SVD and transfering the article\cite{MFLBCB23}, and the Laboratory for Industrial Research (Nissanken) for the finacial aid to the travel expense to INSA in November 
The author is grateful to Prof. S.J. Brodsky, Prof. E. Hitzer and Prof. V. Ariel for helpful communication, 
 Prof. M. Arai and Prof. K. Hamada for allowing the use of a workstation as their research collaborator, and Mr. Wu in the laboratory for helps in Python programmings. Thanks are also due to the library of the Tokyo Institute of Technology and that for mathematical science of the University of Tokyo for allowing consultation of references.}

\vskip 0.5 true cm

\end{document}